\documentclass[twocolumn,showpacs,preprintnumbers,amsmath,amssymb]{revtex4-1}

\usepackage{bbm}
\usepackage{subfigure}
\usepackage{amsmath}
\usepackage{epsfig,psfrag}
\usepackage{dcolumn}
\usepackage{bm}
\usepackage{graphicx}
\usepackage{version}
\usepackage[normalem]{ulem}

\newcommand{\av}[1]{\left<#1\right>}

\newcommand{\PD}{P}
\newcommand{\FD}{F}

\usepackage{times}
\usepackage{color}

\begin{document}
\title{Drude weight fluctuations in many-body localized systems}
\author{Michele Filippone}
\author{Piet W. Brouwer}
\author{Jens Eisert}
\author{Felix von Oppen}
\affiliation{\mbox{Dahlem Center for Complex Quantum Systems and Fachbereich Physik, Freie Universit\"at Berlin, 14195 Berlin, Germany}}

\begin{abstract}
We numerically investigate the distribution of Drude weights $D$ of many-body states in disordered one-dimensional interacting electron systems across the transition to a many-body localized phase. Drude weights are proportional to  the spectral curvatures induced by magnetic fluxes in mesoscopic rings. They offer a method to relate the transition to the many-body localized phase to transport properties. In the delocalized regime, we find that the Drude weight distribution at a fixed disorder configuration agrees well with the random-matrix-theory prediction $P(D) \propto (\gamma^2+D^2)^{-3/2}$, although the distribution width $\gamma$ strongly fluctuates between disorder realizations. A crossover is observed towards a distribution with different large-$D$ asymptotics deep in the many-body localized phase, which however differs from the commonly expected Cauchy distribution.  We show that the average distribution width $\av \gamma$, rescaled by $L\Delta$, $\Delta$ being the average level spacing in the middle of the spectrum and $L$ the systems size, is an efficient probe of the many-body localization transition, as it increases/vanishes exponentially in the delocalized/localized phase.
\end{abstract}


\maketitle

\textit{Introduction.--- }
Electron-electron interactions may drive a disordered electronic system through a delocalization transition at finite temperature \cite{basko06,basko06short,mirlin_mbl}:
Without interactions, Anderson localization implies a vanishing conductivity in one and two dimensions, independent of the disorder strength \cite{anderson58,abrahams79,lagendijk09,abrahams10}. In contrast, in the presence of electron-electron interactions, even in one 
spatial dimension, the conductivity can take a finite value above a critical temperature.
The persistence of localization in the presence of interactions at low temperatures and/or strong disorder is known as many-body localization. Interest in the properties of the many-body localized phase was recently boosted by the demonstration of exotic properties, such as atypical entanglement growth 
logarithmic in time \cite{znidaric08,bardarson12,serbyn13}, anomalous spectral statistics of the many-particle spectrum \cite{oganesyan07,luitz15}, and its connection to equilibration and violation of the eigenstate thermalization hypothesis \cite{goold15,eisert15}. Recently, the first experimental observations 
showing key signatures of a many-body localization transition were reported in systems of cold atoms in optical lattices \cite{schreiber15,bordia15}.

In particular for numerical studies it remains difficult to directly relate the many-body localization transition to the ability of the system to conduct current. The difficulty can be partly attributed to the lack of reliable analytical tools and partly to the relatively small system sizes attainable by numerical approaches.
Recent works in this direction showed substantial modification of dynamic quantities across the transition \cite{lev14,vasseur15,lev15,luitz15b,friesdorf15} as well as atypical behavior of both the stationary \cite{karrasch15} and the finite-frequency conductance \cite{gopalakrishnan15,gopalakrishnan15b}. 

In this letter, we suggest an alternative approach to address the conduction of current across the many-body localization transition, by studying the behavior of the Drude weights $D_n$ of many-body states in one-dimensional interacting disordered systems. The interest in this approach consists in its ability to address stationary transport properties in the presence of both disorder and interactions, without the need to couple the system to source and drain reservoirs. The intuitive idea underlying the approach is that one can distinguish between a metal and an insulator by inspecting the eigenvalue variations under changes of boundary conditions. A magnetic flux $\phi$ in mesoscopic rings is responsible for a twist in the periodic boundary conditions, to which the system responds with persistent currents \cite{buttiker83,kulik10,saminadayar04,bleszynski09}. Drude weights describe the current response to variations of $\phi$ and are related to the curvature of the {\em many-body} eigen-energies $E_n$ \cite{kohn64,shastry90,millis90,fye91,scalapino92,giamarchi88,bouzerar94} (the first derivatives $\partial E_n/\partial \phi$ at $\phi=0$ vanish because of time-reversal symmetry),
\begin{equation}\label{eq:drude}
D_n=\frac L2\left.\frac{\partial^2 E_n}{\partial\phi^2}\right|_{\phi=0}.
\end{equation}
The Drude weights $D_n$ have strong level-to-level fluctuations, so that we must consider their full probability density $\PD$ as a function of $D$. An important argument by Thouless relates the width of this distribution to the average conductance \cite{edwards72,thouless74,akkermans92}. The first derivatives $\partial E_n/\partial \phi$ at a finite flux $\phi$, {\em i.e.}, the persistent currents, were investigated
%
%
for the special case of $N=2$ particles \cite{weinmann1995}, confirming an interaction-induced enhancement of the localization length, a precursor of the interaction-induced delocalization in the many-particle system \cite{shepelyansky1994,imry1995}.

An important reference for the interpretation of our results is on the one hand the prediction of random matrix theory (RMT)  for the distribution of level curvatures in response to a generic perturbation \cite{gaspard90,zakrzewski93,vonoppen94,vonoppen95,fyodorov95,braun97}. This distribution is known and has the exact form \cite{vonoppen94,vonoppen95} 
\begin{equation}\label{eq:p1}
  \PD_{{\rm RMT}}(D)= \frac12\frac{\gamma^2}{(\gamma^2+D^2)^{3/2}}
\end{equation}
for time-reversal symmetric systems. Here, $\gamma>0$ is a parameter setting the width of the distribution. We find that in the delocalized phase the functional form of $\PD(D)$ --- for a specific disorder realization and within an energy window small compared to the width of the many-body spectrum --- is well described by the time-reversal symmetric RMT result. On the other hand, in the many-body localized phase the numerically obtained $\PD(D)$ has a different form, reminiscent of the distribution of {\em single-particle} non-interacting Anderson-localized systems  \cite{braun97,titov97}. The distribution of many-body Drude weights in the many-body localized phase is distinctly different from that of many-body Drude weigths without interactions, however.
For the system sizes that are accessible numerically, the width $\gamma$ of the distribution has strong fluctuations between disorder realisations. It shows a clear exponential decay with system size only in the many-body localized phase. 

\textit{Drude weights, level curvatures, and localization.--- }In the absence of dissipative mechanisms, the Drude weight controls the singularity of the optical conductivity at zero frequency, as $\sigma(\omega)=D\delta(\omega)+\sigma_{\rm reg}(\omega)$ (a numerical study of $\sigma_{\rm reg}$ is carried out in Refs.~\cite{gopalakrishnan15,gopalakrishnan15b}, analytical work is presented in Ref.\ \cite{Prosen}). Since the seminal work by Kohn \cite{kohn64}, the scaling of $D$ with  system size $L$ is a criterion to identify the metal to insulator transition in many body systems \cite{shastry90,millis90}. One has $D\rightarrow 0$ for insulating systems and $D\rightarrow e^2 \rho/m^*$ in the metallic case, $\rho$ being the electron density and $m^*$ a renormalized mass. 

The connection to spectral curvatures \eqref{eq:drude} is readily derived for the model system we consider here, interacting spinless fermions on a one-dimensional ring subject to disorder and  magnetic flux \cite{shastry90}. The Hamiltonian is given by \footnote{The parameters have been defined in such a way that for $\phi=0$ and $t=U=1$ the model maps onto the Heisenberg model $\mathcal H=\sum_i\mathbf S_i\cdot\mathbf S_{i+1}+\sum_i\varepsilon_iS^z_i$, ignoring an overall chemical potential. The Heisenberg model is commonly considered in the literature for many-body localization, see \textit{e.g.} Ref.\ \cite{luitz15}.}
\begin{equation}
\begin{aligned}
\label{eq:ham}
\mathcal H (\phi) =& \mathcal T(\phi) + 
  \sum_{j=1}^L
  \left(
  \varepsilon_j n_j+U n_jn_{j+1}
  \right), \\
\mathcal T(\phi) =& - \frac{1}{2} \sum_{j=1}^{L} \left[ t(\phi) c^\dagger_j c_{j+1} + t(\phi)^* c^\dagger_{j+1} c_{j} \right].
\end{aligned}
\end{equation} 
Here $\phi$ is the magnetic flux, measured in units of the flux quantum $\Phi_0=h/e$, $\mathcal T(\phi)$ is the kinetic energy, with $t(\phi) =t\, e^{2 \pi i \phi/L}$ 
being the complex flux-dependent hopping amplitude, $U$ is the strength of the nearest-neighbor interaction, $\varepsilon_i$ is the on-site disorder potential drawn uniformly from the interval $[-W,W]$, $c_j$ annihilates a particle at site $j$, and $n_j = c_j^{\dagger} c_j$. The ring geometry is realized by identifying $c_1=c_{L+1}$ . 
Without interactions and for weak disorder, $W \ll t$, the localization length of single particle states at the band center is $\xi = c t^2/W^2$ with $c \approx 26.3$ \cite{kappus1981}, measured in units of the lattice spacing, so that the system size $L$ exceeds the localization length $\xi$ for all energies if $W/t \gtrsim 1.3$. With interactions the model (\ref{eq:ham}) is found to display a transition from a many-body spectrum with level repulsion, characteristic of a delocalized phase, to a spectrum without level repulsion. For $U/t=1$ the transition takes place at $W_{\rm c} \approx 3.6 t$  \cite{luitz15}. 
\begin{figure}[t]
\begin{center}
\includegraphics[width=.48\textwidth]{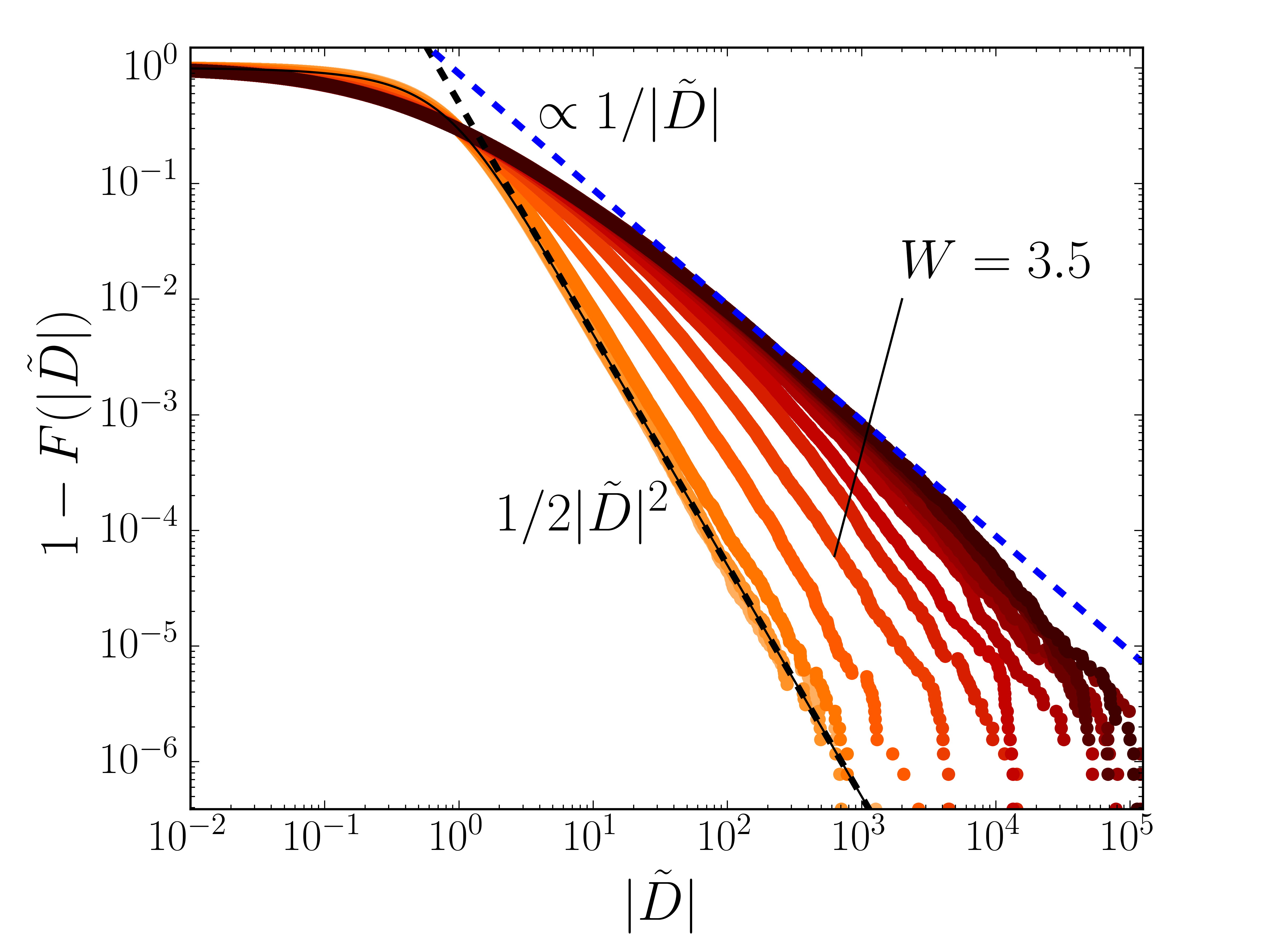}
\caption{Cumulative rescaled Drude weight distribution $\FD$ for disorder strengths $W$ increasing from 1.5 to 7.5 in steps of 0.5 (left to right data series). Each data set is based on 1000 disorder realizations, each contributing 2554 curvatures from states in the middle of the many-body spectrum. The other system parameters are $U/t=1$, $L=16$, and $N=8$ particles. For $W \lesssim 2.5$ the system is in the ergodic phase and the distribution is well approximated by the RMT prediction $\FD_{{\rm RMT}}$ (solid thin line), see Eq.\ (\ref{eq:p1}). Deep in the many-body localized phase ($W/t\gtrsim 5 $), the distribution converges towards a different one with longer tails. }\label{fig:cross}
\end{center}
\end{figure}

For the Hamiltonian \eqref{eq:ham} the current operator reads
\begin{equation}
\mathcal I=\frac{i}{2L}\sum_{j=1}^L\left[t(\phi) c^\dagger_j c_{j+1}- t(\phi)^* c^\dagger_{j+1} c_{j}\right] \nonumber = - \frac{1}{2 \pi} \frac{\partial \mathcal H}{\partial \phi},
\end{equation}
implying that the many-body state vector $|\psi_n\rangle$ of energy $E_n$ carries a persistent current ${\mathcal I}_n = -(1/2 \pi) \partial E_n/\partial \phi$. 
In the vicinity of zero fluxes $\mathcal H$ can be expanded  as $\mathcal H(\phi)=\mathcal H(0)-2\pi\phi\,\mathcal I-2\pi^2\phi^2\mathcal T(0)/L^2+O(\phi^3)$. To second order in $\phi$, the energy shifts read 
$E_n(\phi)-E_n(0)\approx\phi^2 D_n/e^2L$, where the Drude weight of $|\psi_n\rangle$ is given by
\begin{equation}\label{eq:stiff}
D_n=e^2\frac{4\pi^2}L\left[-\frac12\av{\mathcal T}+L^2\sum_{m\neq n}\frac{|\langle \psi_n|\mathcal I|\psi_m\rangle |^2}{E_n-E_m}\right]\,.
\end{equation}
The same expression for the Drude weight can be obtained from the Kubo formula \footnote{Notice that the Drude weights given by Eq. \eqref{eq:stiff} are strongly sensitive to the choice of boundary conditions for finite system sizes \cite{rigol08}. Nevertheless, the possibility to generate finite persistent current $\mathcal I $, whose first derivative in $\phi$  leads directly to Eq. \eqref{eq:drude}, is only possible by assuming periodic boundary conditions. }. 
The assumption of uncorrelated energy levels 
and non-fluctuating matrix elements of the current operator on the one hand  leads to a Cauchy curvature distribution $\PD \propto (\gamma^2 + D^2)^{-1}$ \cite{edwards72,thouless74}. On the other hand, as mentioned in the introduction, a random-matrix  distribution gives the Drude weight distribution of Eq.\ (\ref{eq:p1}) \cite{gaspard90,zakrzewski93,vonoppen94,vonoppen95,fyodorov95,braun97}.

\begin{figure*}[ht!!]
\includegraphics[width=0.325\textwidth]{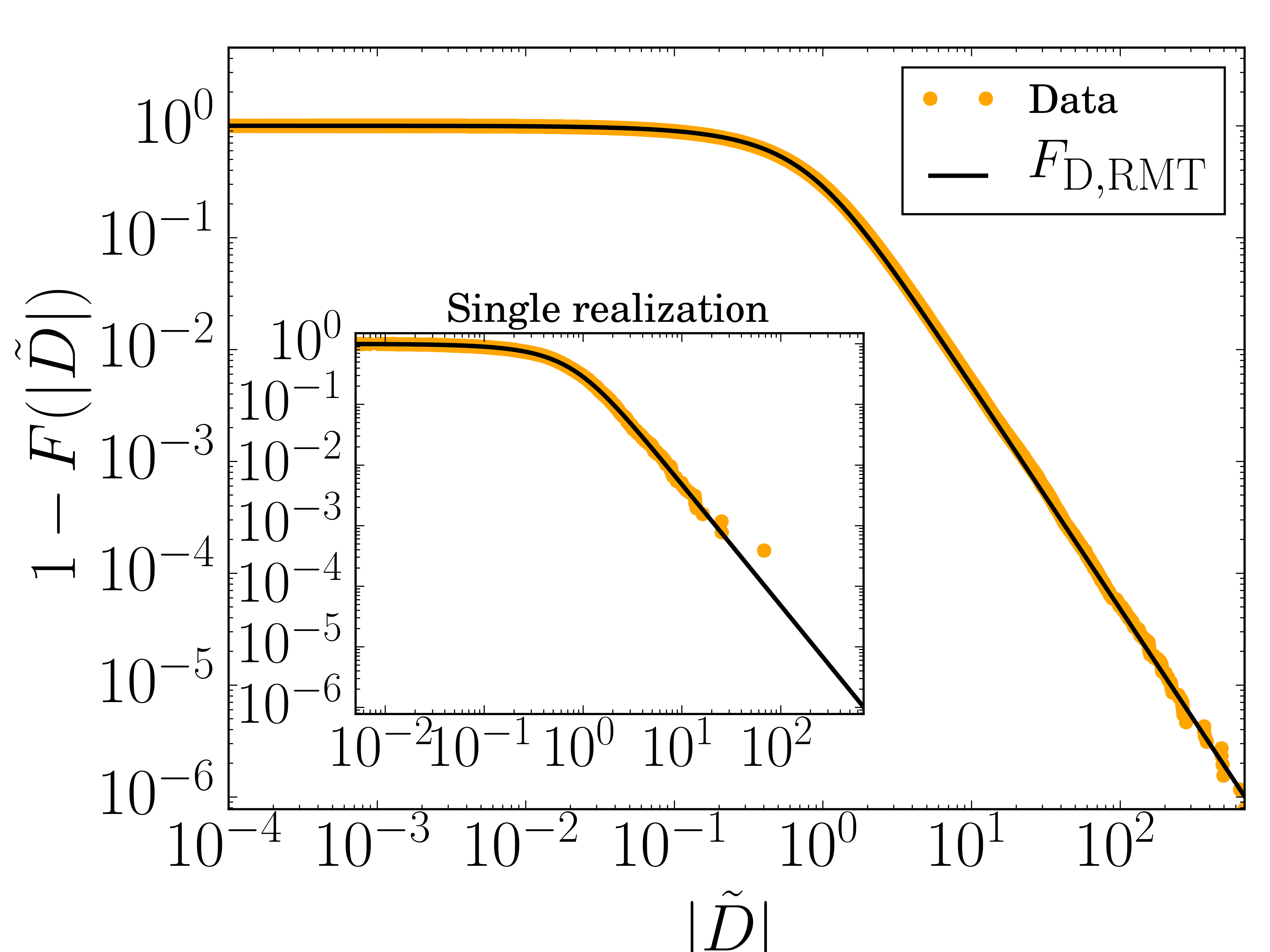}
\includegraphics[width=0.325\textwidth]{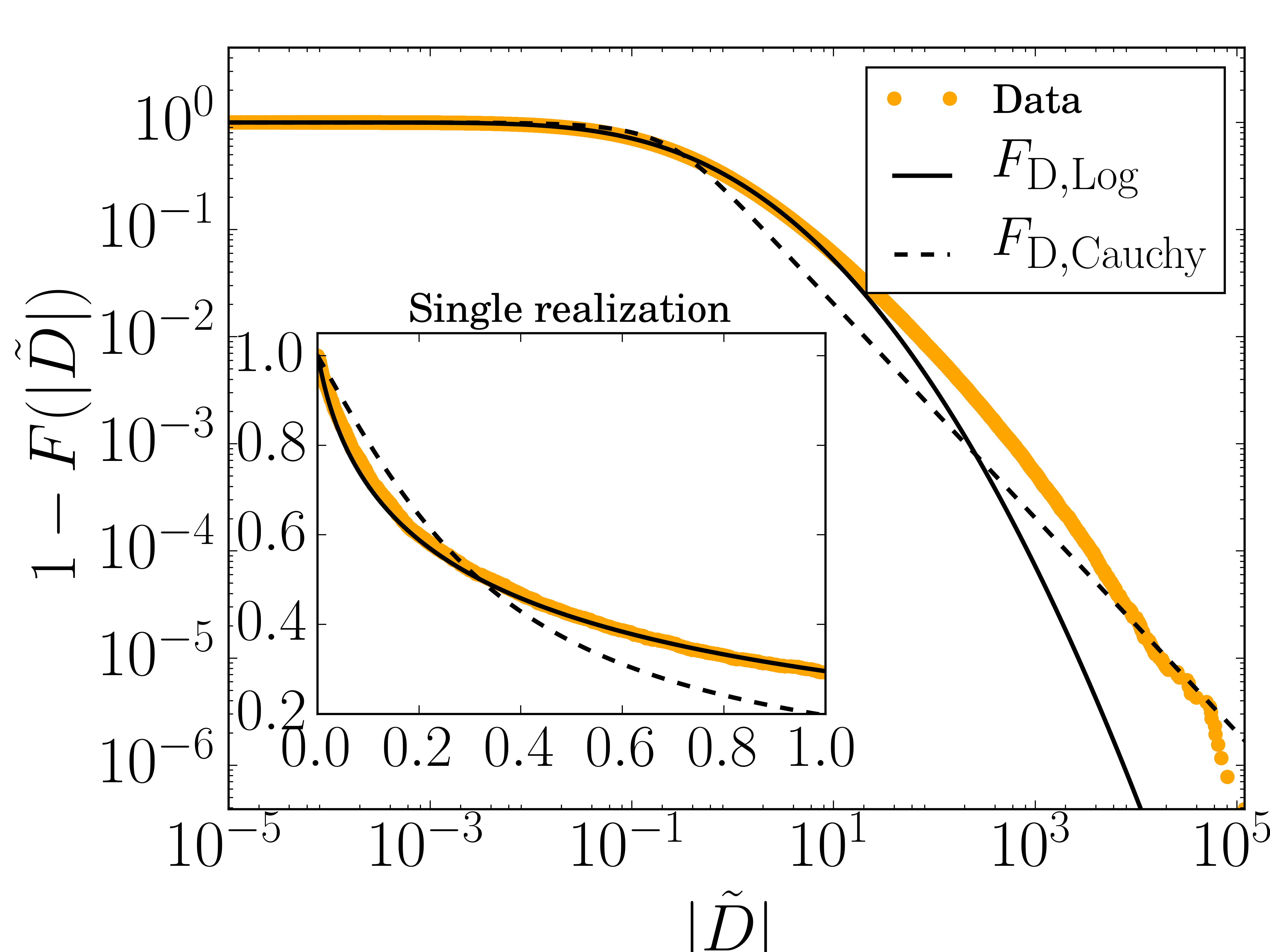}
\includegraphics[width=0.325\textwidth]{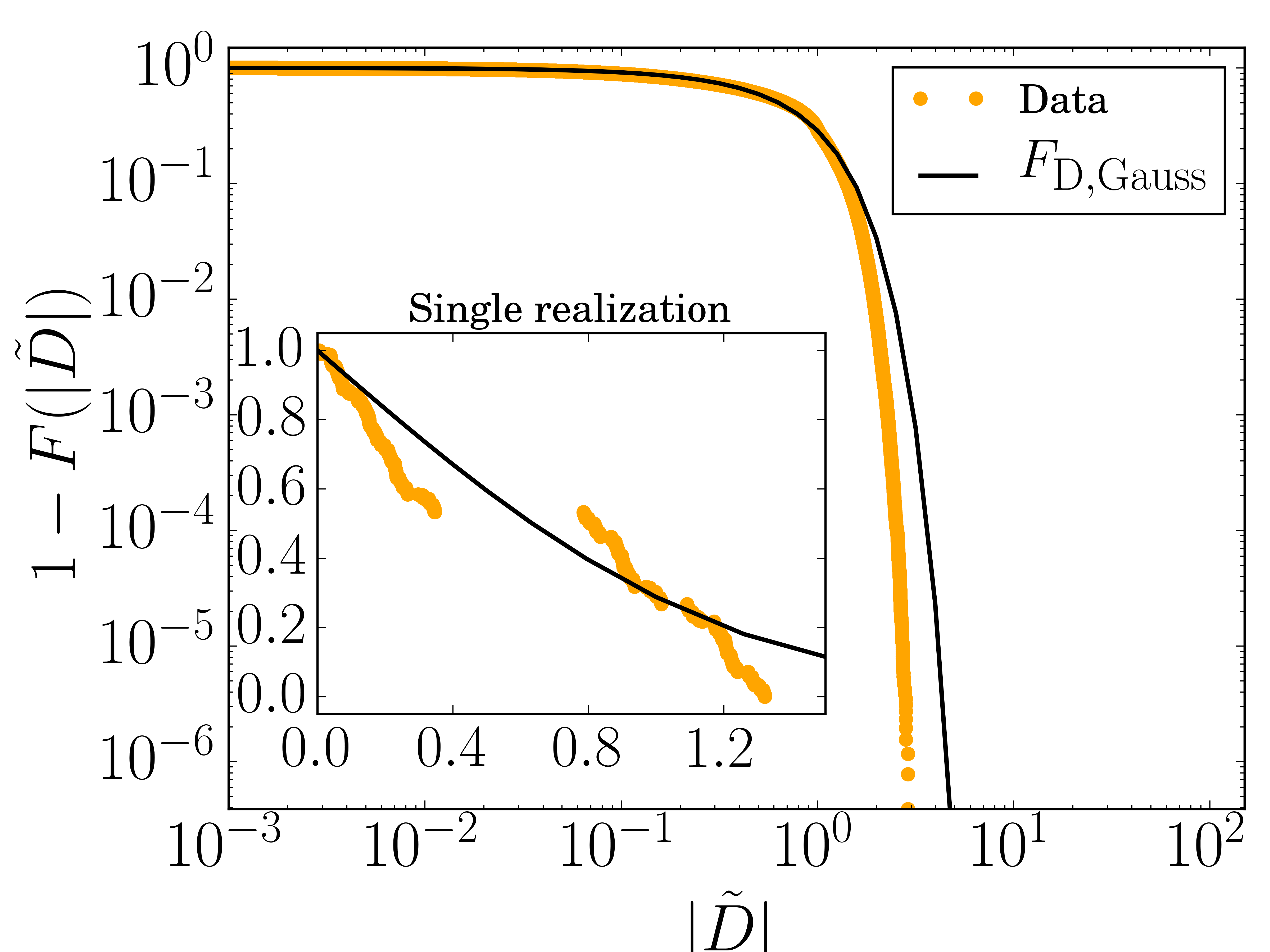}
\caption{Drude-weight distribution for $W/t=2$ (left) and $W/t=5.5$ (center), again at $U/t=1$ for the left and the center figures. The main panels are for rescaled distributions of 2574 states in the middle of the many-body spectrum, averaged over 1000 disorder realizations. The insets are for single disorder realizations. The other parameters are chosen as in Fig.\ \ref{fig:cross}.  For $W/t=2$ both the single-realization and the averaged distributions are in excellent agreement with the RMT prediction. For  $W/t=5.5$ a large part of the distribution is well described by a log-normal distribution, whereas the Cauchy distribution of Refs.\ \cite{edwards72,thouless74,monthus16} does not provide a good fit.  The right panel (main and inset) shows $\FD(D)$ in the absence of interactions, $U=0$.  The curvature distribution for a single realization in the inset shows lack of self-averaging.}\label{fig:distrib}
\end{figure*}   

\textit{Drude weight distribution.---} We have numerically calculated the level curvatures for the full many-body spectrum by exact diagonalization of the Hamiltonian \eqref{eq:ham} at half filling up to $L=16$ sites. We show results for the cumulative distribution function $\FD$  of absolute values $|D|$,
\begin{equation} 
  \FD (D) = \int^{|D|}_{-|D|} dx\,\PD(x).
\end{equation}
To extract a (cumulative) distribution from the numerically obtained level curvatures we consider $M$ many-body levels near the center of the spectrum for a fixed disorder configuration. The exponentially high number of many-body levels ensures that even taking a small fraction of the total many-body spectrum still gives a large number of levels $M$ ($M=2554$ for the center 20\% of many-body levels for $L=16$). To facilitate comparison to the RMT prediction (\ref{eq:p1}), which has $\FD_{{\rm RMT}}(D) = |D|/\sqrt{\gamma^2 + D^2}$, we define the width $\gamma$ of the distribution as that value of $D$ for which $F_{{\rm D}}(D) = \FD_{{\rm RMT}}(\gamma) = 1/\sqrt{2}$.

We find that different disorder realizations with equal strength $W$ give Drude-weight distributions $\PD$ with the same shape, but with different widths. To reduce statistical errors when inspecting the shapes of the distribution functions, we therefore determine the width $\gamma$ of the distribution for each disorder realization separately, rescale the Drude weights $D \mapsto \tilde D = D/\gamma$, such that rescaled distributions have unit width, and then combine distributions from different disorder realizations. Results for such rescaled Drude-weight distributions are shown in Fig.\ \ref{fig:cross}. For $W/t \lesssim 2.5$ the shape of the distribution is in excellent agreement with the RMT prediction \eqref{eq:p1}, see also the left panel of Fig.\ \ref{fig:distrib}. For $W/t \gtrsim 3$ the distribution starts deviating from Eq.\ \eqref{eq:p1}, although the tails (at least initially) continue to scale $\propto |\tilde D|^{-2}$. We attribute the deviation from the RMT prediction for disorder strengths well below $W_{\rm c}$ to finite-size effects, which were also found, in the same way, to cause a ``premature'' transition of the level statistics from random-matrix-like to Poisson, see Ref.\ \cite{luitz15}. For $W \gtrsim W_{\rm c}$, the system enters the many-body localized phase. Because of finite size effects the progression between the ergodic and localized phases appears as a crossover, not as a sharp transition. A crossover of similar width was observed in Ref.\ \cite{luitz15}. Deep in the many-body localized regime ($W/t\gtrsim5$), the distribution converges towards a distribution with significantly longer tails than the RMT distribution (\ref{eq:p1}). Although there is some hint of an intermediate tail scaling $\propto \tilde D^{-1}$, the over-all shape of the distribution in the localized regime is not consistent with the Cauchy distribution of Refs.\ \cite{edwards72,thouless74,monthus16}. A deviation from the Cauchy distribution must be attributed to correlations between the spacings of many-body energy levels and matrix elements of the current operator $\mathcal{I}$, see Eq.\ (\ref{eq:stiff}). Such correlations appear naturally in the localized regime,
taking into account that nearby energy levels generically result from
states ``far apart'' in Fock space, so that matrix elements of (local)
single-particle operators such as $\mathcal{I}$ are strongly suppressed. What is more, while interactions will modify the structure of many-body eigenstates, this is not expected to necessarily lead to a very large deviation in relevant matrix elements, for the same reason that only overlaps reflecting nearby energy levels contribute significantly to the tails of the distribution.

Figure \ref{fig:distrib} shows more detailed results for representative disorder strengths $W/t=2$ and $W/t=5.5$  below and above the many-body localization transition, as well as a comparison with the non-interacting case. The insets show cumulative distributions for a single disorder realization, confirming that our averaging procedure, in the interacting case, does not lead to any systematic deformations of the shape of the distribution function. We also considered different system sizes $L$ (at fixed electron density) \footnote{See Supplemental Material.}, showing that the RMT result Eq. \eqref{eq:p1} is reproduced independently of $L$ in the delocalized phase. This is not the case in the localized regime, in which the distribution tails appear to be sensitive to the system size, although we could not find a tendency towards a Cauchy distribution upon increasing $L$.

Remarkably, for intermediate curvatures, the distribution function in the localized regime is well approximated by a log-normal distribution, see Fig.\ \ref{fig:distrib}, center. 
We note that a log-normal distribution has also been found a good approximation for the curvature distribution of single-particle levels in non-interacting Anderson models with strong disorder \cite{braun97,titov97}. However, this form of the distribution does not necessarily carry over to the many-body curvature distribution for the non-interacting case: Since many-body level curvatures are sums of single-particle level curvatures, it is reasonable to expect that they have a Gaussian distribution as a consequence of the central limit theorem, with non-Gaussian tails to reflect the large fluctuations of the single-particle curvature distribution. Such a distribution is distinctly different from the many-body curvature distribution we observe for the localized phase of the interacting system, see Fig.\ \ref{fig:distrib} (center). For the small system sizes we address here, however, this Gaussian distribution for the non-interacting case has not fully developed yet, see Fig.\ \ref{fig:distrib} (right).

\begin{figure}[t]
\begin{center}
\includegraphics[width=.48\textwidth]{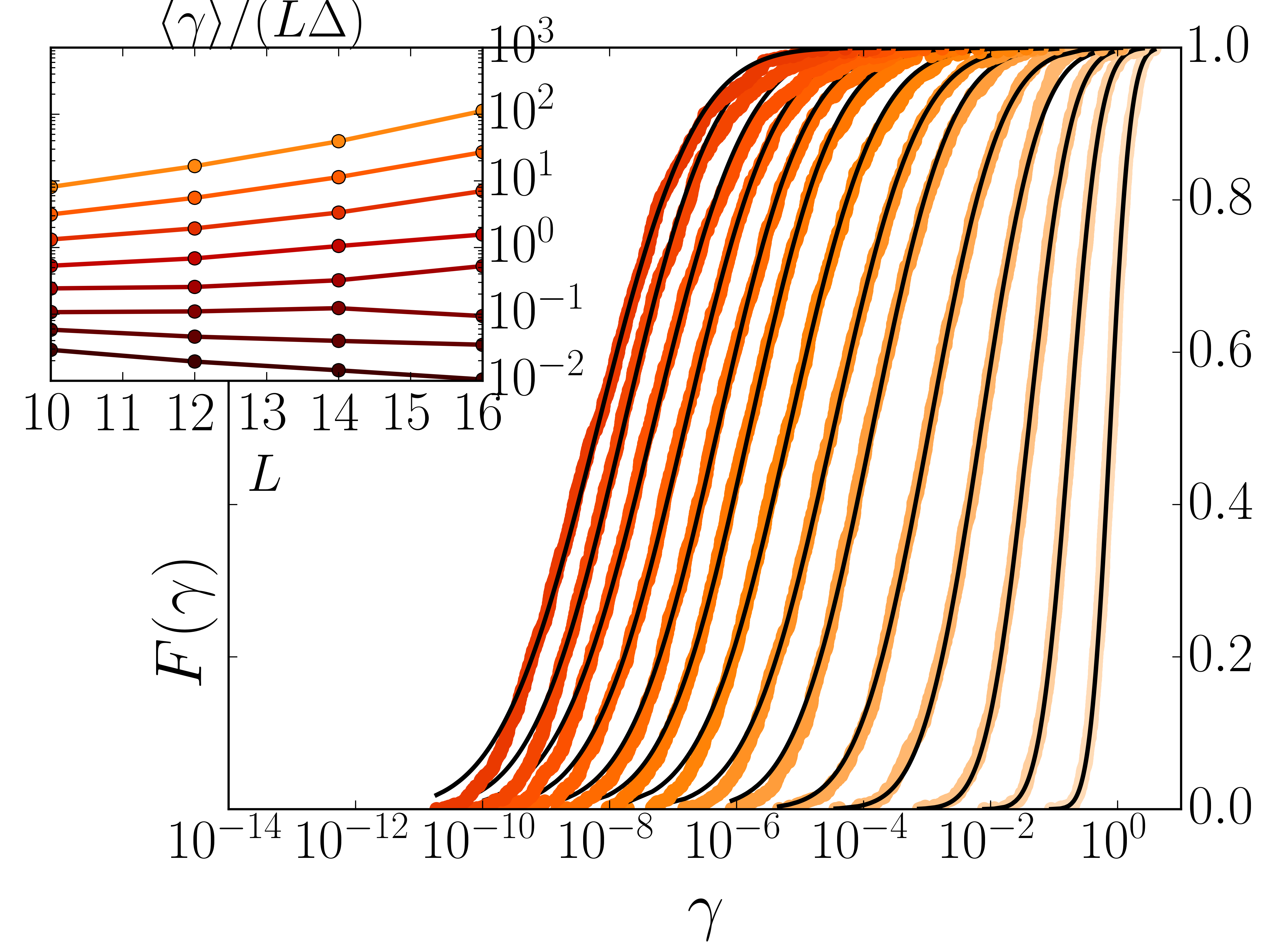}
\caption{Main panel: Cumulative distribution functions of the Drude weight distribution widths  $\gamma$ for disorder strength $W/t$ increasing from $1.5$ to $8$ in steps of $0.5$ (right to left data series). We consider $10^3$ realizations per disorder value. The solid black lines increase a log-normal fit to the data. Parameters of the simulations are as in Fig.\ \ref{fig:cross}. Inset: Average width $\av \gamma$ of the Drude weight distribution (rescaled by $L\Delta$, $\Delta$ being the average level spacing in the middle of the spectrum) vs.\ system size $L$ for disorder strength $W/t$ increasing from $1.5$ to $5$ in steps of $0.5$ (top to bottom data series).}\label{fig:gammi}
\end{center}
\end{figure}

\textit{Fluctuations of the width of the distribution.---} While the shape of the Drude weight distribution was found to be independent of the precise disorder realization, we find that the width $\gamma$ of the distribution has large sample-to-sample fluctuations. Figure \ref{fig:gammi} shows the (cumulative) probability distribution of the widths $\gamma$. This width distribution is well approximated by a log-normal distribution for $W \lesssim W_{\rm c}$, whereas we find that the tails at small (large) $\gamma$ are below (above) log-normal in the many-body localized regime. The average $\av \gamma$ decreases with increasing disorder, whereas the magnitude of the fluctuations increases. This is consistent with the width $\gamma$ being a measure of conductance \cite{edwards72,thouless74}. We attribute the origin of the width fluctuations to finite size effects. Indeed, we find that the variance of the width distribution decreases with system size $L$. However, since the average $\av \gamma$ also decreases with $L$, inset, we cannot settle the question whether the fluctuations of $\gamma$ disappear relative to the average for the limited system sizes attainable in our numerical simulations. The average $\av \gamma$ shows a clear exponential decay in the localized phase; in the delocalized phase we observe a decrease with system size, but could not draw any firm conclusions regarding its functional form. Alternatively, the adimensional quantity $\av\gamma/L\Delta$, in which $\Delta$ is the average level spacing in the middle of the many-body spectrum, is an effective probe of the many-body localization transition. The inset of Fig. \ref{fig:gammi} shows  that this quantity switches from an exponential increase to decrease with system size $L$ across the many-body localization transition, a behavior observed for related quantities in Refs. \cite{serbyn15,monthus16}.

To investigate the strong sample-to-sample fluctuations of the width $\gamma$ of the Drude weight distribution $\PD$, we also calculated the distribution of the matrix elements $\mathcal I_{n,m} = \langle \psi_n | \mathcal I | \psi_m \rangle$ of the current operator (a similar quantity was considered in Ref.~\cite{serbyn15}). Consistently with the large fluctuations of $\gamma$ observed in our numerical simulations, we find large sample-to-sample fluctuations of the mean square $\overline{\mathcal I_{n,m}^2}$, with the average $\overline{\cdots}$ taken with respect to the $M$ many-body state vectors $|\psi_{n,m}\rangle$ within the energy window around the center of the spectrum at a fixed disorder realization. The sample-to-sample fluctuations of $\ln \overline{\mathcal I_{n,m}^2}$ are found to be statistically correlated with the sample-to-sample fluctuations of $\ln \gamma$ (Pearson correlation coefficient $\gtrsim 0.35$, with a slight maximum near the localization transition). The correlations are even stronger, if we consider correlations between $\ln \gamma$ and $\ln \overline{\mathcal I_{n,n+1}^2}$, such that only current matrix elements between neighboring energy levels are included (correlation coefficient $\gtrsim 0.45$), consistent with the expectation based on Eq.\ (\ref{eq:stiff}).


\textit{Conclusions.---} We numerically studied the statistical distribution of Drude weights of many-body states for one-dimensional interacting electrons. We find that the shape of the Drude weight distribution shows clear differences between the weak-disorder and strong-disorder phases, consistent with the onset of a many-body localized phase at strong disorder. The shape of the distribution is still sensitive to the system sizes we could attain in the localized regime. This motivates further studies applying  more refined numerical approaches to address bigger system sizes, in which the study of this issue could be made more definitive. The width $\gamma$ of the Drude weight distribution, defined with respect to a collection of many-body states taken near the center of the spectrum, is commonly associated with the system's conductance. For the relatively small system sizes we could consider, we found large sample-to-sample fluctuations of the distribution width $\gamma$. In the many-body localized phase the disorder average of $\gamma$ shows a clear exponential decay with system size, signaling the suppression of transport. We also showed that the dimensionless quantity $\av \gamma/L\Delta$  discriminates effectively between the localized and delocalized regime. 

Following the seminal ideas of Kohn \cite{kohn64} and Thouless \cite{thouless74}, the sensitivity of the quantum eigenstates to boundary conditions played a crucial role in developing the scaling theory of localization for non-interacting systems \cite{abrahams79}. A thorough understanding of the Drude weights -- measuring the sensitivity of the many-body states to boundary conditions -- may thus contribute an important ingredient to recent attempts \cite{vosk15,potter15,serbyn15} of formulating a scaling theory of the many-body localization transition. A deeper understanding of transport properties may also help in devising novel devices, such  as quantum memories to reliably store quantum information for a long time, stabilized by suitably exploiting many-body localization. It is the hope that the  present work will stimulate such endeavors.

\textit{ Acknowledgments.---} MF heartily thanks Mathis Friesdorf for considerable help at the early stages of this work
and Jens Dreger for assistance. We thank Dmitry Abanin, Jens Bardarson, Daniel Braun, Thierry Giamarchi, Ulrich Krause, David J. Luitz, C\'ecile Monthus, Bj\"orn Sbierski, and Luka Trifunovic for useful discussions. This work was supported by the DFG (CRC/Transregio 183, EI 519/7-1),
the ERC (TAQ), and the EU (AQuS, RAQUEL)

\bibliographystyle{apsrev4-1}
\bibliography{bib_mbl}

\begin{thebibliography}{58}%
\makeatletter
\providecommand \@ifxundefined [1]{%
 \@ifx{#1\undefined}
}%
\providecommand \@ifnum [1]{%
 \ifnum #1\expandafter \@firstoftwo
 \else \expandafter \@secondoftwo
 \fi
}%
\providecommand \@ifx [1]{%
 \ifx #1\expandafter \@firstoftwo
 \else \expandafter \@secondoftwo
 \fi
}%
\providecommand \natexlab [1]{#1}%
\providecommand \enquote  [1]{``#1''}%
\providecommand \bibnamefont  [1]{#1}%
\providecommand \bibfnamefont [1]{#1}%
\providecommand \citenamefont [1]{#1}%
\providecommand \href@noop [0]{\@secondoftwo}%
\providecommand \href [0]{\begingroup \@sanitize@url \@href}%
\providecommand \@href[1]{\@@startlink{#1}\@@href}%
\providecommand \@@href[1]{\endgroup#1\@@endlink}%
\providecommand \@sanitize@url [0]{\catcode `\\12\catcode `\$12\catcode
  `\&12\catcode `\#12\catcode `\^12\catcode `\_12\catcode `\%12\relax}%
\providecommand \@@startlink[1]{}%
\providecommand \@@endlink[0]{}%
\providecommand \url  [0]{\begingroup\@sanitize@url \@url }%
\providecommand \@url [1]{\endgroup\@href {#1}{\urlprefix }}%
\providecommand \urlprefix  [0]{URL }%
\providecommand \Eprint [0]{\href }%
\providecommand \doibase [0]{http://dx.doi.org/}%
\providecommand \selectlanguage [0]{\@gobble}%
\providecommand \bibinfo  [0]{\@secondoftwo}%
\providecommand \bibfield  [0]{\@secondoftwo}%
\providecommand \translation [1]{[#1]}%
\providecommand \BibitemOpen [0]{}%
\providecommand \bibitemStop [0]{}%
\providecommand \bibitemNoStop [0]{.\EOS\space}%
\providecommand \EOS [0]{\spacefactor3000\relax}%
\providecommand \BibitemShut  [1]{\csname bibitem#1\endcsname}%
\let\auto@bib@innerbib\@empty
\bibitem [{\citenamefont {Basko}\ \emph {et~al.}(2006)\citenamefont {Basko},
  \citenamefont {Aleiner},\ and\ \citenamefont {Altshuler}}]{basko06}%
  \BibitemOpen
  \bibfield  {author} {\bibinfo {author} {\bibfnamefont {D.~M.}\ \bibnamefont
  {Basko}}, \bibinfo {author} {\bibfnamefont {I.~L.}\ \bibnamefont {Aleiner}},
  \ and\ \bibinfo {author} {\bibfnamefont {B.~L.}\ \bibnamefont {Altshuler}},\
  }\href@noop {} {\bibfield  {journal} {\bibinfo  {journal} {Ann.\ Phys.}\
  }\textbf {\bibinfo {volume} {321}},\ \bibinfo {pages} {1126} (\bibinfo {year}
  {2006})}\BibitemShut {NoStop}%
\bibitem [{\citenamefont {Basko}\ \emph {et~al.}()\citenamefont {Basko},
  \citenamefont {Aleiner},\ and\ \citenamefont {Altshuler}}]{basko06short}%
  \BibitemOpen
  \bibfield  {author} {\bibinfo {author} {\bibfnamefont {D.~M.}\ \bibnamefont
  {Basko}}, \bibinfo {author} {\bibfnamefont {I.~L.}\ \bibnamefont {Aleiner}},
  \ and\ \bibinfo {author} {\bibfnamefont {B.~L.}\ \bibnamefont {Altshuler}},\
  }\href@noop {} {\bibinfo  {journal} {in {\em Problems of Condensed Matter
  Physics}, eds.\ A.\ L.\ Ivanov and S\ G.\ Tikhodeev (Oxford Univ.\ Press.,
  Oxford, 2007)}\ ,\ \bibinfo {pages} {p.\ 50}}\BibitemShut {NoStop}%
\bibitem [{\citenamefont {Gornyi}\ \emph {et~al.}(2005)\citenamefont {Gornyi},
  \citenamefont {Mirlin},\ and\ \citenamefont {Polyakov}}]{mirlin_mbl}%
  \BibitemOpen
\bibfield  {journal} {  }\bibfield  {author} {\bibinfo {author} {\bibfnamefont
  {I.~V.}\ \bibnamefont {Gornyi}}, \bibinfo {author} {\bibfnamefont {A.~D.}\
  \bibnamefont {Mirlin}}, \ and\ \bibinfo {author} {\bibfnamefont {D.~G.}\
  \bibnamefont {Polyakov}},\ }\href@noop {} {\bibfield  {journal} {\bibinfo
  {journal} {Phys. Rev. Lett.}\ }\textbf {\bibinfo {volume} {95}},\ \bibinfo
  {pages} {206603} (\bibinfo {year} {2005})}\BibitemShut {NoStop}%
\bibitem [{\citenamefont {Anderson}(1958)}]{anderson58}%
  \BibitemOpen
  \bibfield  {author} {\bibinfo {author} {\bibfnamefont {P.~W.}\ \bibnamefont
  {Anderson}},\ }\href {\doibase 10.1103/PhysRev.109.1492} {\bibfield
  {journal} {\bibinfo  {journal} {Phys. Rev.}\ }\textbf {\bibinfo {volume}
  {109}},\ \bibinfo {pages} {1492} (\bibinfo {year} {1958})}\BibitemShut
  {NoStop}%
\bibitem [{\citenamefont {Abrahams}\ \emph {et~al.}(1979)\citenamefont
  {Abrahams}, \citenamefont {Anderson}, \citenamefont {Licciardello},\ and\
  \citenamefont {Ramakrishnan}}]{abrahams79}%
  \BibitemOpen
  \bibfield  {author} {\bibinfo {author} {\bibfnamefont {E.}~\bibnamefont
  {Abrahams}}, \bibinfo {author} {\bibfnamefont {P.~W.}\ \bibnamefont
  {Anderson}}, \bibinfo {author} {\bibfnamefont {D.~C.}\ \bibnamefont
  {Licciardello}}, \ and\ \bibinfo {author} {\bibfnamefont {T.~V.}\
  \bibnamefont {Ramakrishnan}},\ }\href {\doibase 10.1103/PhysRevLett.42.673}
  {\bibfield  {journal} {\bibinfo  {journal} {Phys. Rev. Lett.}\ }\textbf
  {\bibinfo {volume} {42}},\ \bibinfo {pages} {673} (\bibinfo {year}
  {1979})}\BibitemShut {NoStop}%
\bibitem [{\citenamefont {Lagendijk}\ \emph {et~al.}(2009)\citenamefont
  {Lagendijk}, \citenamefont {van Tiggelen},\ and\ \citenamefont
  {Wiersma}}]{lagendijk09}%
  \BibitemOpen
  \bibfield  {author} {\bibinfo {author} {\bibfnamefont {A.}~\bibnamefont
  {Lagendijk}}, \bibinfo {author} {\bibfnamefont {B.}~\bibnamefont {van
  Tiggelen}}, \ and\ \bibinfo {author} {\bibfnamefont {D.~S.}\ \bibnamefont
  {Wiersma}},\ }\href@noop {} {\bibfield  {journal} {\bibinfo  {journal} {Phys.
  Today}\ }\textbf {\bibinfo {volume} {62}},\ \bibinfo {pages} {24} (\bibinfo
  {year} {2009})}\BibitemShut {NoStop}%
\bibitem [{\citenamefont {Abrahams}(2010)}]{abrahams10}%
  \BibitemOpen
  \bibfield  {author} {\bibinfo {author} {\bibfnamefont {E.}~\bibnamefont
  {Abrahams}},\ }\href@noop {} {\emph {\bibinfo {title} {50 Years of Anderson
  Localization}}},\ Vol.~\bibinfo {volume} {24}\ (\bibinfo  {publisher} {World
  Scientific},\ \bibinfo {year} {2010})\BibitemShut {NoStop}%
\bibitem [{\citenamefont {\ifmmode \check{Z}\else
  \v{Z}\fi{}nidari\ifmmode~\check{c}\else \v{c}\fi{}}\ \emph
  {et~al.}(2008)\citenamefont {\ifmmode \check{Z}\else
  \v{Z}\fi{}nidari\ifmmode~\check{c}\else \v{c}\fi{}}, \citenamefont {Prosen},\
  and\ \citenamefont {Prelov\ifmmode~\check{s}\else
  \v{s}\fi{}ek}}]{znidaric08}%
  \BibitemOpen
  \bibfield  {author} {\bibinfo {author} {\bibfnamefont {M.}~\bibnamefont
  {\ifmmode \check{Z}\else \v{Z}\fi{}nidari\ifmmode~\check{c}\else
  \v{c}\fi{}}}, \bibinfo {author} {\bibfnamefont {T.}~\bibnamefont {Prosen}}, \
  and\ \bibinfo {author} {\bibfnamefont {P.}~\bibnamefont
  {Prelov\ifmmode~\check{s}\else \v{s}\fi{}ek}},\ }\href {\doibase
  10.1103/PhysRevB.77.064426} {\bibfield  {journal} {\bibinfo  {journal} {Phys.
  Rev. B}\ }\textbf {\bibinfo {volume} {77}},\ \bibinfo {pages} {064426}
  (\bibinfo {year} {2008})}\BibitemShut {NoStop}%
\bibitem [{\citenamefont {Bardarson}\ \emph {et~al.}(2012)\citenamefont
  {Bardarson}, \citenamefont {Pollmann},\ and\ \citenamefont
  {Moore}}]{bardarson12}%
  \BibitemOpen
  \bibfield  {author} {\bibinfo {author} {\bibfnamefont {J.~H.}\ \bibnamefont
  {Bardarson}}, \bibinfo {author} {\bibfnamefont {F.}~\bibnamefont {Pollmann}},
  \ and\ \bibinfo {author} {\bibfnamefont {J.~E.}\ \bibnamefont {Moore}},\
  }\href {\doibase 10.1103/PhysRevLett.109.017202} {\bibfield  {journal}
  {\bibinfo  {journal} {Phys. Rev. Lett.}\ }\textbf {\bibinfo {volume} {109}},\
  \bibinfo {pages} {017202} (\bibinfo {year} {2012})}\BibitemShut {NoStop}%
\bibitem [{\citenamefont {Serbyn}\ \emph {et~al.}(2013)\citenamefont {Serbyn},
  \citenamefont {Papi\ifmmode~\acute{c}\else \'{c}\fi{}},\ and\ \citenamefont
  {Abanin}}]{serbyn13}%
  \BibitemOpen
  \bibfield  {author} {\bibinfo {author} {\bibfnamefont {M.}~\bibnamefont
  {Serbyn}}, \bibinfo {author} {\bibfnamefont {Z.}~\bibnamefont
  {Papi\ifmmode~\acute{c}\else \'{c}\fi{}}}, \ and\ \bibinfo {author}
  {\bibfnamefont {D.~A.}\ \bibnamefont {Abanin}},\ }\href {\doibase
  10.1103/PhysRevLett.110.260601} {\bibfield  {journal} {\bibinfo  {journal}
  {Phys. Rev. Lett.}\ }\textbf {\bibinfo {volume} {110}},\ \bibinfo {pages}
  {260601} (\bibinfo {year} {2013})}\BibitemShut {NoStop}%
\bibitem [{\citenamefont {Oganesyan}\ and\ \citenamefont
  {Huse}(2007)}]{oganesyan07}%
  \BibitemOpen
  \bibfield  {author} {\bibinfo {author} {\bibfnamefont {V.}~\bibnamefont
  {Oganesyan}}\ and\ \bibinfo {author} {\bibfnamefont {D.~A.}\ \bibnamefont
  {Huse}},\ }\href {\doibase 10.1103/PhysRevB.75.155111} {\bibfield  {journal}
  {\bibinfo  {journal} {Phys. Rev. B}\ }\textbf {\bibinfo {volume} {75}},\
  \bibinfo {pages} {155111} (\bibinfo {year} {2007})}\BibitemShut {NoStop}%
\bibitem [{\citenamefont {Luitz}\ \emph {et~al.}(2015)\citenamefont {Luitz},
  \citenamefont {Laflorencie},\ and\ \citenamefont {Alet}}]{luitz15}%
  \BibitemOpen
  \bibfield  {author} {\bibinfo {author} {\bibfnamefont {D.~J.}\ \bibnamefont
  {Luitz}}, \bibinfo {author} {\bibfnamefont {N.}~\bibnamefont {Laflorencie}},
  \ and\ \bibinfo {author} {\bibfnamefont {F.}~\bibnamefont {Alet}},\ }\href
  {\doibase 10.1103/PhysRevB.91.081103} {\bibfield  {journal} {\bibinfo
  {journal} {Phys. Rev. B}\ }\textbf {\bibinfo {volume} {91}},\ \bibinfo
  {pages} {081103} (\bibinfo {year} {2015})}\BibitemShut {NoStop}%
\bibitem [{\citenamefont {Goold}\ \emph {et~al.}(2015)\citenamefont {Goold},
  \citenamefont {Gogolin}, \citenamefont {Clark}, \citenamefont {Eisert},
  \citenamefont {Scardicchio},\ and\ \citenamefont {Silva}}]{goold15}%
  \BibitemOpen
  \bibfield  {author} {\bibinfo {author} {\bibfnamefont {J.}~\bibnamefont
  {Goold}}, \bibinfo {author} {\bibfnamefont {C.}~\bibnamefont {Gogolin}},
  \bibinfo {author} {\bibfnamefont {S.~R.}\ \bibnamefont {Clark}}, \bibinfo
  {author} {\bibfnamefont {J.}~\bibnamefont {Eisert}}, \bibinfo {author}
  {\bibfnamefont {A.}~\bibnamefont {Scardicchio}}, \ and\ \bibinfo {author}
  {\bibfnamefont {A.}~\bibnamefont {Silva}},\ }\href {\doibase
  10.1103/PhysRevB.92.180202} {\bibfield  {journal} {\bibinfo  {journal} {Phys.
  Rev. B}\ }\textbf {\bibinfo {volume} {92}},\ \bibinfo {pages} {180202}
  (\bibinfo {year} {2015})}\BibitemShut {NoStop}%
\bibitem [{\citenamefont {Eisert}\ \emph {et~al.}(2015)\citenamefont {Eisert},
  \citenamefont {Friesdorf},\ and\ \citenamefont {Gogolin}}]{eisert15}%
  \BibitemOpen
  \bibfield  {author} {\bibinfo {author} {\bibfnamefont {J.}~\bibnamefont
  {Eisert}}, \bibinfo {author} {\bibfnamefont {M.}~\bibnamefont {Friesdorf}}, \
  and\ \bibinfo {author} {\bibfnamefont {C.}~\bibnamefont {Gogolin}},\
  }\href@noop {} {\bibfield  {journal} {\bibinfo  {journal} {Nature Phys.}\
  }\textbf {\bibinfo {volume} {11}},\ \bibinfo {pages} {124} (\bibinfo {year}
  {2015})}\BibitemShut {NoStop}%
\bibitem [{\citenamefont {Schreiber}\ \emph {et~al.}(2015)\citenamefont
  {Schreiber}, \citenamefont {Hodgman}, \citenamefont {Bordia}, \citenamefont
  {L{\"u}schen}, \citenamefont {Fischer}, \citenamefont {Vosk}, \citenamefont
  {Altman}, \citenamefont {Schneider},\ and\ \citenamefont
  {Bloch}}]{schreiber15}%
  \BibitemOpen
  \bibfield  {author} {\bibinfo {author} {\bibfnamefont {M.}~\bibnamefont
  {Schreiber}}, \bibinfo {author} {\bibfnamefont {S.~S.}\ \bibnamefont
  {Hodgman}}, \bibinfo {author} {\bibfnamefont {P.}~\bibnamefont {Bordia}},
  \bibinfo {author} {\bibfnamefont {H.~P.}\ \bibnamefont {L{\"u}schen}},
  \bibinfo {author} {\bibfnamefont {M.~H.}\ \bibnamefont {Fischer}}, \bibinfo
  {author} {\bibfnamefont {R.}~\bibnamefont {Vosk}}, \bibinfo {author}
  {\bibfnamefont {E.}~\bibnamefont {Altman}}, \bibinfo {author} {\bibfnamefont
  {U.}~\bibnamefont {Schneider}}, \ and\ \bibinfo {author} {\bibfnamefont
  {I.}~\bibnamefont {Bloch}},\ }\href@noop {} {\bibfield  {journal} {\bibinfo
  {journal} {Science}\ }\textbf {\bibinfo {volume} {349}},\ \bibinfo {pages}
  {842} (\bibinfo {year} {2015})}\BibitemShut {NoStop}%
\bibitem [{\citenamefont {Bordia}\ \emph {et~al.}(2016)\citenamefont {Bordia},
  \citenamefont {L{\"u}schen}, \citenamefont {Hodgman}, \citenamefont
  {Schreiber}, \citenamefont {Bloch},\ and\ \citenamefont
  {Schneider}}]{bordia15}%
  \BibitemOpen
  \bibfield  {author} {\bibinfo {author} {\bibfnamefont {P.}~\bibnamefont
  {Bordia}}, \bibinfo {author} {\bibfnamefont {H.~P.}\ \bibnamefont
  {L{\"u}schen}}, \bibinfo {author} {\bibfnamefont {S.~S.}\ \bibnamefont
  {Hodgman}}, \bibinfo {author} {\bibfnamefont {M.}~\bibnamefont {Schreiber}},
  \bibinfo {author} {\bibfnamefont {I.}~\bibnamefont {Bloch}}, \ and\ \bibinfo
  {author} {\bibfnamefont {U.}~\bibnamefont {Schneider}},\ }\href@noop {}
  {\bibfield  {journal} {\bibinfo  {journal} {Phys. Rev. B}\ }\textbf {\bibinfo
  {volume} {116}},\ \bibinfo {pages} {140401} (\bibinfo {year}
  {2016})}\BibitemShut {NoStop}%
\bibitem [{\citenamefont {Bar~Lev}\ and\ \citenamefont
  {Reichman}(2014)}]{lev14}%
  \BibitemOpen
  \bibfield  {author} {\bibinfo {author} {\bibfnamefont {Y.}~\bibnamefont
  {Bar~Lev}}\ and\ \bibinfo {author} {\bibfnamefont {D.~R.}\ \bibnamefont
  {Reichman}},\ }\href {\doibase 10.1103/PhysRevB.89.220201} {\bibfield
  {journal} {\bibinfo  {journal} {Phys. Rev. B}\ }\textbf {\bibinfo {volume}
  {89}},\ \bibinfo {pages} {220201} (\bibinfo {year} {2014})}\BibitemShut
  {NoStop}%
\bibitem [{\citenamefont {Vasseur}\ \emph {et~al.}(2015)\citenamefont
  {Vasseur}, \citenamefont {Parameswaran},\ and\ \citenamefont
  {Moore}}]{vasseur15}%
  \BibitemOpen
  \bibfield  {author} {\bibinfo {author} {\bibfnamefont {R.}~\bibnamefont
  {Vasseur}}, \bibinfo {author} {\bibfnamefont {S.~A.}\ \bibnamefont
  {Parameswaran}}, \ and\ \bibinfo {author} {\bibfnamefont {J.~E.}\
  \bibnamefont {Moore}},\ }\href {\doibase 10.1103/PhysRevB.91.140202}
  {\bibfield  {journal} {\bibinfo  {journal} {Phys. Rev. B}\ }\textbf {\bibinfo
  {volume} {91}},\ \bibinfo {pages} {140202} (\bibinfo {year}
  {2015})}\BibitemShut {NoStop}%
\bibitem [{\citenamefont {Bar~Lev}\ \emph {et~al.}(2015)\citenamefont
  {Bar~Lev}, \citenamefont {Cohen},\ and\ \citenamefont {Reichman}}]{lev15}%
  \BibitemOpen
  \bibfield  {author} {\bibinfo {author} {\bibfnamefont {Y.}~\bibnamefont
  {Bar~Lev}}, \bibinfo {author} {\bibfnamefont {G.}~\bibnamefont {Cohen}}, \
  and\ \bibinfo {author} {\bibfnamefont {D.~R.}\ \bibnamefont {Reichman}},\
  }\href {\doibase 10.1103/PhysRevLett.114.100601} {\bibfield  {journal}
  {\bibinfo  {journal} {Phys. Rev. Lett.}\ }\textbf {\bibinfo {volume} {114}},\
  \bibinfo {pages} {100601} (\bibinfo {year} {2015})}\BibitemShut {NoStop}%
\bibitem [{\citenamefont {Luitz}\ \emph {et~al.}(2016)\citenamefont {Luitz},
  \citenamefont {Laflorencie},\ and\ \citenamefont {Alet}}]{luitz15b}%
  \BibitemOpen
  \bibfield  {author} {\bibinfo {author} {\bibfnamefont {D.~J.}\ \bibnamefont
  {Luitz}}, \bibinfo {author} {\bibfnamefont {N.}~\bibnamefont {Laflorencie}},
  \ and\ \bibinfo {author} {\bibfnamefont {F.}~\bibnamefont {Alet}},\
  }\href@noop {} {\bibfield  {journal} {\bibinfo  {journal} {Phys. Rev. B}\
  }\textbf {\bibinfo {volume} {93}},\ \bibinfo {pages} {060201(R)} (\bibinfo
  {year} {2016})}\BibitemShut {NoStop}%
\bibitem [{\citenamefont {Friesdorf}\ \emph {et~al.}(2015)\citenamefont
  {Friesdorf}, \citenamefont {Werner}, \citenamefont {Brown}, \citenamefont
  {Scholz},\ and\ \citenamefont {Eisert}}]{friesdorf15}%
  \BibitemOpen
  \bibfield  {author} {\bibinfo {author} {\bibfnamefont {M.}~\bibnamefont
  {Friesdorf}}, \bibinfo {author} {\bibfnamefont {A.~H.}\ \bibnamefont
  {Werner}}, \bibinfo {author} {\bibfnamefont {W.}~\bibnamefont {Brown}},
  \bibinfo {author} {\bibfnamefont {V.~B.}\ \bibnamefont {Scholz}}, \ and\
  \bibinfo {author} {\bibfnamefont {J.}~\bibnamefont {Eisert}},\ }\href
  {\doibase 10.1103/PhysRevLett.114.170505} {\bibfield  {journal} {\bibinfo
  {journal} {Phys. Rev. Lett.}\ }\textbf {\bibinfo {volume} {114}},\ \bibinfo
  {pages} {170505} (\bibinfo {year} {2015})}\BibitemShut {NoStop}%
\bibitem [{\citenamefont {Karrasch}\ and\ \citenamefont
  {Moore}(2015)}]{karrasch15}%
  \BibitemOpen
  \bibfield  {author} {\bibinfo {author} {\bibfnamefont {C.}~\bibnamefont
  {Karrasch}}\ and\ \bibinfo {author} {\bibfnamefont {J.~E.}\ \bibnamefont
  {Moore}},\ }\href {\doibase 10.1103/PhysRevB.92.115108} {\bibfield  {journal}
  {\bibinfo  {journal} {Phys. Rev. B}\ }\textbf {\bibinfo {volume} {92}},\
  \bibinfo {pages} {115108} (\bibinfo {year} {2015})}\BibitemShut {NoStop}%
\bibitem [{\citenamefont {Gopalakrishnan}\ \emph {et~al.}(2015)\citenamefont
  {Gopalakrishnan}, \citenamefont {M\"uller}, \citenamefont {Khemani},
  \citenamefont {Knap}, \citenamefont {Demler},\ and\ \citenamefont
  {Huse}}]{gopalakrishnan15}%
  \BibitemOpen
  \bibfield  {author} {\bibinfo {author} {\bibfnamefont {S.}~\bibnamefont
  {Gopalakrishnan}}, \bibinfo {author} {\bibfnamefont {M.}~\bibnamefont
  {M\"uller}}, \bibinfo {author} {\bibfnamefont {V.}~\bibnamefont {Khemani}},
  \bibinfo {author} {\bibfnamefont {M.}~\bibnamefont {Knap}}, \bibinfo {author}
  {\bibfnamefont {E.}~\bibnamefont {Demler}}, \ and\ \bibinfo {author}
  {\bibfnamefont {D.~A.}\ \bibnamefont {Huse}},\ }\href {\doibase
  10.1103/PhysRevB.92.104202} {\bibfield  {journal} {\bibinfo  {journal} {Phys.
  Rev. B}\ }\textbf {\bibinfo {volume} {92}},\ \bibinfo {pages} {104202}
  (\bibinfo {year} {2015})}\BibitemShut {NoStop}%
\bibitem [{\citenamefont {Gopalakrishnan}\ \emph {et~al.}(2016)\citenamefont
  {Gopalakrishnan}, \citenamefont {Agarwal}, \citenamefont {Huse},
  \citenamefont {Demler},\ and\ \citenamefont {Knap}}]{gopalakrishnan15b}%
  \BibitemOpen
  \bibfield  {author} {\bibinfo {author} {\bibfnamefont {S.}~\bibnamefont
  {Gopalakrishnan}}, \bibinfo {author} {\bibfnamefont {K.}~\bibnamefont
  {Agarwal}}, \bibinfo {author} {\bibfnamefont {D.~A.}\ \bibnamefont {Huse}},
  \bibinfo {author} {\bibfnamefont {E.}~\bibnamefont {Demler}}, \ and\ \bibinfo
  {author} {\bibfnamefont {M.}~\bibnamefont {Knap}},\ }\href@noop {} {\bibfield
   {journal} {\bibinfo  {journal} {Phys. Rev. B}\ }\textbf {\bibinfo {volume}
  {93}},\ \bibinfo {pages} {134206} (\bibinfo {year} {2016})}\BibitemShut
  {NoStop}%
\bibitem [{\citenamefont {B{\"u}ttiker}\ \emph {et~al.}(1983)\citenamefont
  {B{\"u}ttiker}, \citenamefont {Imry},\ and\ \citenamefont
  {Landauer}}]{buttiker83}%
  \BibitemOpen
  \bibfield  {author} {\bibinfo {author} {\bibfnamefont {M.}~\bibnamefont
  {B{\"u}ttiker}}, \bibinfo {author} {\bibfnamefont {Y.}~\bibnamefont {Imry}},
  \ and\ \bibinfo {author} {\bibfnamefont {R.}~\bibnamefont {Landauer}},\
  }\href@noop {} {\bibfield  {journal} {\bibinfo  {journal} {Physics Lett. A}\
  }\textbf {\bibinfo {volume} {96}},\ \bibinfo {pages} {365} (\bibinfo {year}
  {1983})}\BibitemShut {NoStop}%
\bibitem [{\citenamefont {Kulik}(2010)}]{kulik10}%
  \BibitemOpen
  \bibfield  {author} {\bibinfo {author} {\bibfnamefont {I.~O.}\ \bibnamefont
  {Kulik}},\ }\href@noop {} {\bibfield  {journal} {\bibinfo  {journal} {Low
  Temp. Phys.}\ }\textbf {\bibinfo {volume} {36}},\ \bibinfo {pages} {841}
  (\bibinfo {year} {2010})}\BibitemShut {NoStop}%
\bibitem [{\citenamefont {Saminadayar}\ \emph {et~al.}(2004)\citenamefont
  {Saminadayar}, \citenamefont {B{\"a}uerle},\ and\ \citenamefont
  {Mailly}}]{saminadayar04}%
  \BibitemOpen
  \bibfield  {author} {\bibinfo {author} {\bibfnamefont {L.}~\bibnamefont
  {Saminadayar}}, \bibinfo {author} {\bibfnamefont {C.}~\bibnamefont
  {B{\"a}uerle}}, \ and\ \bibinfo {author} {\bibfnamefont {D.}~\bibnamefont
  {Mailly}},\ }\href@noop {} {\bibfield  {journal} {\bibinfo  {journal} {in
  {\em Encyclopedia of Nanoscience and Nanotechnology}, eds. H.S. Nalwa,
  Valencia, CA, American Scientific}\ }\textbf {\bibinfo {volume} {3}},\
  \bibinfo {pages} {267} (\bibinfo {year} {2004})}\BibitemShut {NoStop}%
\bibitem [{\citenamefont {Bleszynski-Jayich}\ \emph {et~al.}(2009)\citenamefont
  {Bleszynski-Jayich}, \citenamefont {Shanks}, \citenamefont {Peaudecerf},
  \citenamefont {Ginossar}, \citenamefont {von Oppen}, \citenamefont
  {Glazman},\ and\ \citenamefont {Harris}}]{bleszynski09}%
  \BibitemOpen
  \bibfield  {author} {\bibinfo {author} {\bibfnamefont {A.}~\bibnamefont
  {Bleszynski-Jayich}}, \bibinfo {author} {\bibfnamefont {W.}~\bibnamefont
  {Shanks}}, \bibinfo {author} {\bibfnamefont {B.}~\bibnamefont {Peaudecerf}},
  \bibinfo {author} {\bibfnamefont {E.}~\bibnamefont {Ginossar}}, \bibinfo
  {author} {\bibfnamefont {F.}~\bibnamefont {von Oppen}}, \bibinfo {author}
  {\bibfnamefont {L.}~\bibnamefont {Glazman}}, \ and\ \bibinfo {author}
  {\bibfnamefont {J.}~\bibnamefont {Harris}},\ }\href@noop {} {\bibfield
  {journal} {\bibinfo  {journal} {Science}\ }\textbf {\bibinfo {volume}
  {326}},\ \bibinfo {pages} {272} (\bibinfo {year} {2009})}\BibitemShut
  {NoStop}%
\bibitem [{\citenamefont {Kohn}(1964)}]{kohn64}%
  \BibitemOpen
  \bibfield  {author} {\bibinfo {author} {\bibfnamefont {W.}~\bibnamefont
  {Kohn}},\ }\href {\doibase 10.1103/PhysRev.133.A171} {\bibfield  {journal}
  {\bibinfo  {journal} {Phys. Rev.}\ }\textbf {\bibinfo {volume} {133}},\
  \bibinfo {pages} {A171} (\bibinfo {year} {1964})}\BibitemShut {NoStop}%
\bibitem [{\citenamefont {Shastry}\ and\ \citenamefont
  {Sutherland}(1990)}]{shastry90}%
  \BibitemOpen
  \bibfield  {author} {\bibinfo {author} {\bibfnamefont {B.~S.}\ \bibnamefont
  {Shastry}}\ and\ \bibinfo {author} {\bibfnamefont {B.}~\bibnamefont
  {Sutherland}},\ }\href {\doibase 10.1103/PhysRevLett.65.243} {\bibfield
  {journal} {\bibinfo  {journal} {Phys. Rev. Lett.}\ }\textbf {\bibinfo
  {volume} {65}},\ \bibinfo {pages} {243} (\bibinfo {year} {1990})}\BibitemShut
  {NoStop}%
\bibitem [{\citenamefont {Millis}\ and\ \citenamefont
  {Coppersmith}(1990)}]{millis90}%
  \BibitemOpen
  \bibfield  {author} {\bibinfo {author} {\bibfnamefont {A.~J.}\ \bibnamefont
  {Millis}}\ and\ \bibinfo {author} {\bibfnamefont {S.~N.}\ \bibnamefont
  {Coppersmith}},\ }\href {\doibase 10.1103/PhysRevB.42.10807} {\bibfield
  {journal} {\bibinfo  {journal} {Phys. Rev. B}\ }\textbf {\bibinfo {volume}
  {42}},\ \bibinfo {pages} {10807} (\bibinfo {year} {1990})}\BibitemShut
  {NoStop}%
\bibitem [{\citenamefont {Fye}\ \emph {et~al.}(1991)\citenamefont {Fye},
  \citenamefont {Martins}, \citenamefont {Scalapino}, \citenamefont {Wagner},\
  and\ \citenamefont {Hanke}}]{fye91}%
  \BibitemOpen
  \bibfield  {author} {\bibinfo {author} {\bibfnamefont {R.~M.}\ \bibnamefont
  {Fye}}, \bibinfo {author} {\bibfnamefont {M.~J.}\ \bibnamefont {Martins}},
  \bibinfo {author} {\bibfnamefont {D.~J.}\ \bibnamefont {Scalapino}}, \bibinfo
  {author} {\bibfnamefont {J.}~\bibnamefont {Wagner}}, \ and\ \bibinfo {author}
  {\bibfnamefont {W.}~\bibnamefont {Hanke}},\ }\href {\doibase
  10.1103/PhysRevB.44.6909} {\bibfield  {journal} {\bibinfo  {journal} {Phys.
  Rev. B}\ }\textbf {\bibinfo {volume} {44}},\ \bibinfo {pages} {6909}
  (\bibinfo {year} {1991})}\BibitemShut {NoStop}%
\bibitem [{\citenamefont {Scalapino}\ \emph {et~al.}(1992)\citenamefont
  {Scalapino}, \citenamefont {White},\ and\ \citenamefont
  {Zhang}}]{scalapino92}%
  \BibitemOpen
  \bibfield  {author} {\bibinfo {author} {\bibfnamefont {D.~J.}\ \bibnamefont
  {Scalapino}}, \bibinfo {author} {\bibfnamefont {S.~R.}\ \bibnamefont
  {White}}, \ and\ \bibinfo {author} {\bibfnamefont {S.~C.}\ \bibnamefont
  {Zhang}},\ }\href {\doibase 10.1103/PhysRevLett.68.2830} {\bibfield
  {journal} {\bibinfo  {journal} {Phys. Rev. Lett.}\ }\textbf {\bibinfo
  {volume} {68}},\ \bibinfo {pages} {2830} (\bibinfo {year}
  {1992})}\BibitemShut {NoStop}%
\bibitem [{\citenamefont {Giamarchi}\ and\ \citenamefont
  {Schulz}(1988)}]{giamarchi88}%
  \BibitemOpen
  \bibfield  {author} {\bibinfo {author} {\bibfnamefont {T.}~\bibnamefont
  {Giamarchi}}\ and\ \bibinfo {author} {\bibfnamefont {H.~J.}\ \bibnamefont
  {Schulz}},\ }\href {\doibase 10.1103/PhysRevB.37.325} {\bibfield  {journal}
  {\bibinfo  {journal} {Phys. Rev. B}\ }\textbf {\bibinfo {volume} {37}},\
  \bibinfo {pages} {325} (\bibinfo {year} {1988})}\BibitemShut {NoStop}%
\bibitem [{\citenamefont {Bouzerar}\ \emph {et~al.}(1994)\citenamefont
  {Bouzerar}, \citenamefont {Poilblanc},\ and\ \citenamefont
  {Montambaux}}]{bouzerar94}%
  \BibitemOpen
  \bibfield  {author} {\bibinfo {author} {\bibfnamefont {G.}~\bibnamefont
  {Bouzerar}}, \bibinfo {author} {\bibfnamefont {D.}~\bibnamefont {Poilblanc}},
  \ and\ \bibinfo {author} {\bibfnamefont {G.}~\bibnamefont {Montambaux}},\
  }\href {\doibase 10.1103/PhysRevB.49.8258} {\bibfield  {journal} {\bibinfo
  {journal} {Phys. Rev. B}\ }\textbf {\bibinfo {volume} {49}},\ \bibinfo
  {pages} {8258} (\bibinfo {year} {1994})}\BibitemShut {NoStop}%
\bibitem [{\citenamefont {Edwards}\ and\ \citenamefont
  {Thouless}(1972)}]{edwards72}%
  \BibitemOpen
  \bibfield  {author} {\bibinfo {author} {\bibfnamefont {J.}~\bibnamefont
  {Edwards}}\ and\ \bibinfo {author} {\bibfnamefont {D.}~\bibnamefont
  {Thouless}},\ }\href@noop {} {\bibfield  {journal} {\bibinfo  {journal} {J.
  Phys. C: Sol. State Phys.}\ }\textbf {\bibinfo {volume} {5}},\ \bibinfo
  {pages} {807} (\bibinfo {year} {1972})}\BibitemShut {NoStop}%
\bibitem [{\citenamefont {Thouless}(1974)}]{thouless74}%
  \BibitemOpen
  \bibfield  {author} {\bibinfo {author} {\bibfnamefont {D.~J.}\ \bibnamefont
  {Thouless}},\ }\href@noop {} {\bibfield  {journal} {\bibinfo  {journal}
  {Phys. Rep.}\ }\textbf {\bibinfo {volume} {13}},\ \bibinfo {pages} {93}
  (\bibinfo {year} {1974})}\BibitemShut {NoStop}%
\bibitem [{\citenamefont {Akkermans}\ and\ \citenamefont
  {Montambaux}(1992)}]{akkermans92}%
  \BibitemOpen
  \bibfield  {author} {\bibinfo {author} {\bibfnamefont {E.}~\bibnamefont
  {Akkermans}}\ and\ \bibinfo {author} {\bibfnamefont {G.}~\bibnamefont
  {Montambaux}},\ }\href {\doibase 10.1103/PhysRevLett.68.642} {\bibfield
  {journal} {\bibinfo  {journal} {Phys. Rev. Lett.}\ }\textbf {\bibinfo
  {volume} {68}},\ \bibinfo {pages} {642} (\bibinfo {year} {1992})}\BibitemShut
  {NoStop}%
\bibitem [{\citenamefont {Weinmann}\ \emph {et~al.}(1995)\citenamefont
  {Weinmann}, \citenamefont {M\"uller-Groeling}, \citenamefont {Pichard},\ and\
  \citenamefont {Frahm}}]{weinmann1995}%
  \BibitemOpen
  \bibfield  {author} {\bibinfo {author} {\bibfnamefont {D.}~\bibnamefont
  {Weinmann}}, \bibinfo {author} {\bibfnamefont {A.}~\bibnamefont
  {M\"uller-Groeling}}, \bibinfo {author} {\bibfnamefont {J.-L.}\ \bibnamefont
  {Pichard}}, \ and\ \bibinfo {author} {\bibfnamefont {K.}~\bibnamefont
  {Frahm}},\ }\href {\doibase 10.1103/PhysRevLett.75.1598} {\bibfield
  {journal} {\bibinfo  {journal} {Phys. Rev. Lett.}\ }\textbf {\bibinfo
  {volume} {75}},\ \bibinfo {pages} {1598} (\bibinfo {year}
  {1995})}\BibitemShut {NoStop}%
\bibitem [{\citenamefont {Shepelyansky}(1994)}]{shepelyansky1994}%
  \BibitemOpen
  \bibfield  {author} {\bibinfo {author} {\bibfnamefont {D.~L.}\ \bibnamefont
  {Shepelyansky}},\ }\href {\doibase 10.1103/PhysRevLett.73.2607} {\bibfield
  {journal} {\bibinfo  {journal} {Phys. Rev. Lett.}\ }\textbf {\bibinfo
  {volume} {73}},\ \bibinfo {pages} {2607} (\bibinfo {year}
  {1994})}\BibitemShut {NoStop}%
\bibitem [{\citenamefont {Imry}(1995)}]{imry1995}%
  \BibitemOpen
  \bibfield  {author} {\bibinfo {author} {\bibfnamefont {Y.}~\bibnamefont
  {Imry}},\ }\href {http://stacks.iop.org/0295-5075/30/i=7/a=005} {\bibfield
  {journal} {\bibinfo  {journal} {Europhys. Lett.}\ }\textbf {\bibinfo {volume}
  {30}},\ \bibinfo {pages} {405} (\bibinfo {year} {1995})}\BibitemShut
  {NoStop}%
\bibitem [{\citenamefont {Gaspard}\ \emph {et~al.}(1990)\citenamefont
  {Gaspard}, \citenamefont {Rice}, \citenamefont {Mikeska},\ and\ \citenamefont
  {Nakamura}}]{gaspard90}%
  \BibitemOpen
  \bibfield  {author} {\bibinfo {author} {\bibfnamefont {P.}~\bibnamefont
  {Gaspard}}, \bibinfo {author} {\bibfnamefont {S.~A.}\ \bibnamefont {Rice}},
  \bibinfo {author} {\bibfnamefont {H.~J.}\ \bibnamefont {Mikeska}}, \ and\
  \bibinfo {author} {\bibfnamefont {K.}~\bibnamefont {Nakamura}},\ }\href
  {\doibase 10.1103/PhysRevA.42.4015} {\bibfield  {journal} {\bibinfo
  {journal} {Phys. Rev. A}\ }\textbf {\bibinfo {volume} {42}},\ \bibinfo
  {pages} {4015} (\bibinfo {year} {1990})}\BibitemShut {NoStop}%
\bibitem [{\citenamefont {Zakrzewski}\ and\ \citenamefont
  {Delande}(1993)}]{zakrzewski93}%
  \BibitemOpen
  \bibfield  {author} {\bibinfo {author} {\bibfnamefont {J.}~\bibnamefont
  {Zakrzewski}}\ and\ \bibinfo {author} {\bibfnamefont {D.}~\bibnamefont
  {Delande}},\ }\href {\doibase 10.1103/PhysRevE.47.1650} {\bibfield  {journal}
  {\bibinfo  {journal} {Phys. Rev. E}\ }\textbf {\bibinfo {volume} {47}},\
  \bibinfo {pages} {1650} (\bibinfo {year} {1993})}\BibitemShut {NoStop}%
\bibitem [{\citenamefont {von Oppen}(1994)}]{vonoppen94}%
  \BibitemOpen
  \bibfield  {author} {\bibinfo {author} {\bibfnamefont {F.}~\bibnamefont {von
  Oppen}},\ }\href {\doibase 10.1103/PhysRevLett.73.798} {\bibfield  {journal}
  {\bibinfo  {journal} {Phys. Rev. Lett.}\ }\textbf {\bibinfo {volume} {73}},\
  \bibinfo {pages} {798} (\bibinfo {year} {1994})}\BibitemShut {NoStop}%
\bibitem [{\citenamefont {von Oppen}(1995)}]{vonoppen95}%
  \BibitemOpen
  \bibfield  {author} {\bibinfo {author} {\bibfnamefont {F.}~\bibnamefont {von
  Oppen}},\ }\href {\doibase 10.1103/PhysRevE.51.2647} {\bibfield  {journal}
  {\bibinfo  {journal} {Phys. Rev. E}\ }\textbf {\bibinfo {volume} {51}},\
  \bibinfo {pages} {2647} (\bibinfo {year} {1995})}\BibitemShut {NoStop}%
\bibitem [{\citenamefont {Fyodorov}\ and\ \citenamefont
  {Sommers}(1995)}]{fyodorov95}%
  \BibitemOpen
  \bibfield  {author} {\bibinfo {author} {\bibfnamefont {Y.~V.}\ \bibnamefont
  {Fyodorov}}\ and\ \bibinfo {author} {\bibfnamefont {H.-J.}\ \bibnamefont
  {Sommers}},\ }\href {\doibase 10.1103/PhysRevE.51.R2719} {\bibfield
  {journal} {\bibinfo  {journal} {Phys. Rev. E}\ }\textbf {\bibinfo {volume}
  {51}},\ \bibinfo {pages} {R2719} (\bibinfo {year} {1995})}\BibitemShut
  {NoStop}%
\bibitem [{\citenamefont {Braun}\ \emph {et~al.}(1997)\citenamefont {Braun},
  \citenamefont {Hofstetter}, \citenamefont {MacKinnon},\ and\ \citenamefont
  {Montambaux}}]{braun97}%
  \BibitemOpen
  \bibfield  {author} {\bibinfo {author} {\bibfnamefont {D.}~\bibnamefont
  {Braun}}, \bibinfo {author} {\bibfnamefont {E.}~\bibnamefont {Hofstetter}},
  \bibinfo {author} {\bibfnamefont {A.}~\bibnamefont {MacKinnon}}, \ and\
  \bibinfo {author} {\bibfnamefont {G.}~\bibnamefont {Montambaux}},\ }\href
  {\doibase 10.1103/PhysRevB.55.7557} {\bibfield  {journal} {\bibinfo
  {journal} {Phys. Rev. B}\ }\textbf {\bibinfo {volume} {55}},\ \bibinfo
  {pages} {7557} (\bibinfo {year} {1997})}\BibitemShut {NoStop}%
\bibitem [{\citenamefont {Titov}\ \emph {et~al.}(1997)\citenamefont {Titov},
  \citenamefont {Braun},\ and\ \citenamefont {Fyodorov}}]{titov97}%
  \BibitemOpen
  \bibfield  {author} {\bibinfo {author} {\bibfnamefont {M.}~\bibnamefont
  {Titov}}, \bibinfo {author} {\bibfnamefont {D.}~\bibnamefont {Braun}}, \ and\
  \bibinfo {author} {\bibfnamefont {Y.~V.}\ \bibnamefont {Fyodorov}},\
  }\href@noop {} {\bibfield  {journal} {\bibinfo  {journal} {J. Phys. A}\
  }\textbf {\bibinfo {volume} {30}},\ \bibinfo {pages} {L339} (\bibinfo {year}
  {1997})}\BibitemShut {NoStop}%
\bibitem [{\citenamefont {Ilievski}\ and\ \citenamefont
  {Prosen}(2013)}]{Prosen}%
  \BibitemOpen
  \bibfield  {author} {\bibinfo {author} {\bibfnamefont {E.}~\bibnamefont
  {Ilievski}}\ and\ \bibinfo {author} {\bibfnamefont {T.}~\bibnamefont
  {Prosen}},\ }\href@noop {} {\bibfield  {journal} {\bibinfo  {journal}
  {Commun. Math. Phys.}\ }\textbf {\bibinfo {volume} {318}},\ \bibinfo {pages}
  {809} (\bibinfo {year} {2013})}\BibitemShut {NoStop}%
\bibitem [{Note1()}]{Note1}%
  \BibitemOpen
  \bibinfo {note} {The parameters have been defined in such a way that for
  $\phi =0$ and $t=U=1$ the model maps onto the Heisenberg model $\protect
  \mathcal H=\DOTSB \sum@ \slimits@ _i\protect \mathbf S_i\cdot \protect
  \mathbf S_{i+1}+\DOTSB \sum@ \slimits@ _i\varepsilon _iS^z_i$, ignoring an
  overall chemical potential. The Heisenberg model is commonly considered in
  the literature for many-body localization, see \protect \textit {e.g.} Ref.\
  \cite {luitz15}.}\BibitemShut {Stop}%
\bibitem [{\citenamefont {Kappus}\ and\ \citenamefont
  {Wegner}(1981)}]{kappus1981}%
  \BibitemOpen
  \bibfield  {author} {\bibinfo {author} {\bibfnamefont {M.}~\bibnamefont
  {Kappus}}\ and\ \bibinfo {author} {\bibfnamefont {F.}~\bibnamefont
  {Wegner}},\ }\href@noop {} {\bibfield  {journal} {\bibinfo  {journal} {Z.
  Phys. B}\ }\textbf {\bibinfo {volume} {45}},\ \bibinfo {pages} {15} (\bibinfo
  {year} {1981})}\BibitemShut {NoStop}%
\bibitem [{Note2()}]{Note2}%
  \BibitemOpen
  \bibinfo {note} {Notice that the Drude weights given by Eq. \protect \textup
  {\hbox {\mathsurround \z@ \protect \normalfont (\ignorespaces \ref
  {eq:stiff}\unskip \@@italiccorr )}} are strongly sensitive to the choice of
  boundary conditions for finite system sizes \cite {rigol08}. Nevertheless,
  the possibility to generate finite persistent current $\protect \mathcal I $,
  whose first derivative in $\phi $ leads directly to Eq. \protect \textup
  {\hbox {\mathsurround \z@ \protect \normalfont (\ignorespaces \ref
  {eq:drude}\unskip \@@italiccorr )}}, is only possible by assuming periodic
  boundary conditions.}\BibitemShut {Stop}%
\bibitem [{\citenamefont {Monthus}(2016)}]{monthus16}%
  \BibitemOpen
  \bibfield  {author} {\bibinfo {author} {\bibfnamefont {C.}~\bibnamefont
  {Monthus}},\ }\href@noop {} {\bibfield  {journal} {\bibinfo  {journal} {arXiv
  preprint arXiv:1607.00750}\ } (\bibinfo {year} {2016})}\BibitemShut {NoStop}%
\bibitem [{Note3()}]{Note3}%
  \BibitemOpen
  \bibinfo {note} {See Supplemental Material.}\BibitemShut {Stop}%
\bibitem [{\citenamefont {Serbyn}\ \emph {et~al.}(2015)\citenamefont {Serbyn},
  \citenamefont {Papi\ifmmode~\acute{c}\else \'{c}\fi{}},\ and\ \citenamefont
  {Abanin}}]{serbyn15}%
  \BibitemOpen
  \bibfield  {author} {\bibinfo {author} {\bibfnamefont {M.}~\bibnamefont
  {Serbyn}}, \bibinfo {author} {\bibfnamefont {Z.}~\bibnamefont
  {Papi\ifmmode~\acute{c}\else \'{c}\fi{}}}, \ and\ \bibinfo {author}
  {\bibfnamefont {D.~A.}\ \bibnamefont {Abanin}},\ }\href {\doibase
  10.1103/PhysRevX.5.041047} {\bibfield  {journal} {\bibinfo  {journal} {Phys.
  Rev. X}\ }\textbf {\bibinfo {volume} {5}},\ \bibinfo {pages} {041047}
  (\bibinfo {year} {2015})}\BibitemShut {NoStop}%
\bibitem [{\citenamefont {Vosk}\ \emph {et~al.}(2015)\citenamefont {Vosk},
  \citenamefont {Huse},\ and\ \citenamefont {Altman}}]{vosk15}%
  \BibitemOpen
  \bibfield  {author} {\bibinfo {author} {\bibfnamefont {R.}~\bibnamefont
  {Vosk}}, \bibinfo {author} {\bibfnamefont {D.~A.}\ \bibnamefont {Huse}}, \
  and\ \bibinfo {author} {\bibfnamefont {E.}~\bibnamefont {Altman}},\ }\href
  {\doibase 10.1103/PhysRevX.5.031032} {\bibfield  {journal} {\bibinfo
  {journal} {Phys. Rev. X}\ }\textbf {\bibinfo {volume} {5}},\ \bibinfo {pages}
  {031032} (\bibinfo {year} {2015})}\BibitemShut {NoStop}%
\bibitem [{\citenamefont {Potter}\ \emph {et~al.}(2015)\citenamefont {Potter},
  \citenamefont {Vasseur},\ and\ \citenamefont {Parameswaran}}]{potter15}%
  \BibitemOpen
  \bibfield  {author} {\bibinfo {author} {\bibfnamefont {A.~C.}\ \bibnamefont
  {Potter}}, \bibinfo {author} {\bibfnamefont {R.}~\bibnamefont {Vasseur}}, \
  and\ \bibinfo {author} {\bibfnamefont {S.~A.}\ \bibnamefont {Parameswaran}},\
  }\href {\doibase 10.1103/PhysRevX.5.031033} {\bibfield  {journal} {\bibinfo
  {journal} {Phys. Rev. X}\ }\textbf {\bibinfo {volume} {5}},\ \bibinfo {pages}
  {031033} (\bibinfo {year} {2015})}\BibitemShut {NoStop}%
\bibitem [{\citenamefont {Rigol}\ and\ \citenamefont
  {Shastry}(2008)}]{rigol08}%
  \BibitemOpen
  \bibfield  {author} {\bibinfo {author} {\bibfnamefont {M.}~\bibnamefont
  {Rigol}}\ and\ \bibinfo {author} {\bibfnamefont {B.~S.}\ \bibnamefont
  {Shastry}},\ }\href {\doibase 10.1103/PhysRevB.77.161101} {\bibfield
  {journal} {\bibinfo  {journal} {Phys. Rev. B}\ }\textbf {\bibinfo {volume}
  {77}},\ \bibinfo {pages} {161101} (\bibinfo {year} {2008})}\BibitemShut
  {NoStop}%
\end{thebibliography}%

\newpage 


\onecolumngrid

\section*{Supplemental Material for ``Drude weight fluctuations in many-body localized systems''}

\begin{quote}
In this Supplemental Material, we provide detailed finite-size scaling analysis of the Drude weight distribution both in the delocalized and localized regime. We show that finite-size effects are absent in the delocalized regime, while they are strong in the many-body localized phase. 
\end{quote}

\renewcommand{\thesection}{S-\Roman{section}}
\renewcommand{\theequation}{S-\arabic{equation}}
\renewcommand{\thefigure}{S-\arabic{figure}}

\maketitle

\setcounter{figure}{0}

\begin{figure}[h]
\begin{center}
\includegraphics[width=.45\textwidth]{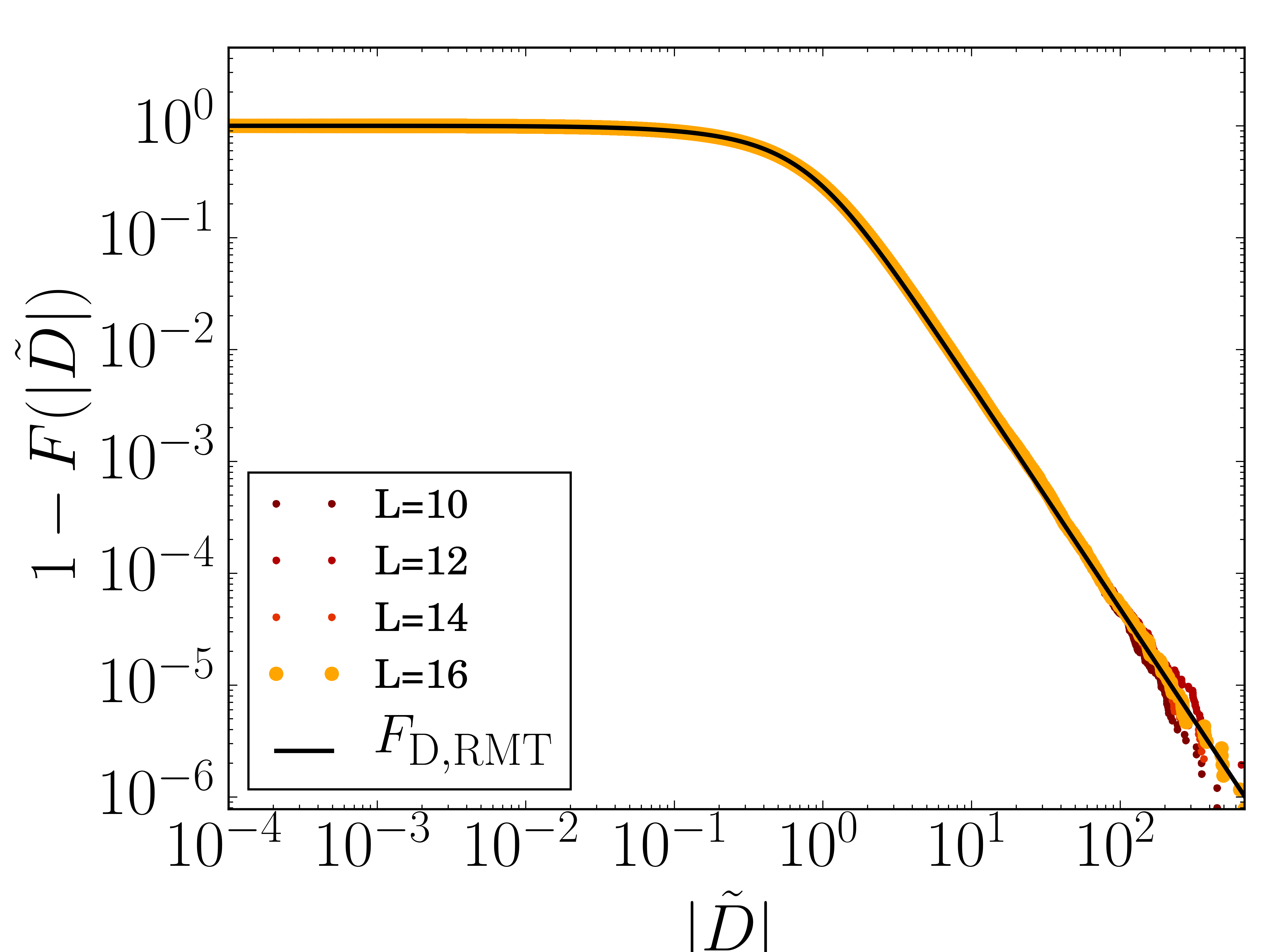}
\caption{Finite-size scaling of the Drude weight distribution in the delocalized regime ($W/t=2.0$). The plot is the same as the left panel in Fig. 2 in the main text, but presents data for different systems sizes, always at half-filling. We consider 50000, 14000, 4000 and 1000 realizations (each contributing with 50, 185, 685 and 2574 states from the middle of the many-body spectrum) for $L=10,12,14$ and 16 respectively. All numerical data series overlap, showing the absence of any size dependence in the delocalized phase, and are in perfect agreement with the RMT prediction (solid black line). }\label{fig:sizermt}
\end{center}
\end{figure}

\begin{figure}[h]
\begin{center}
\includegraphics[width=.45\textwidth]{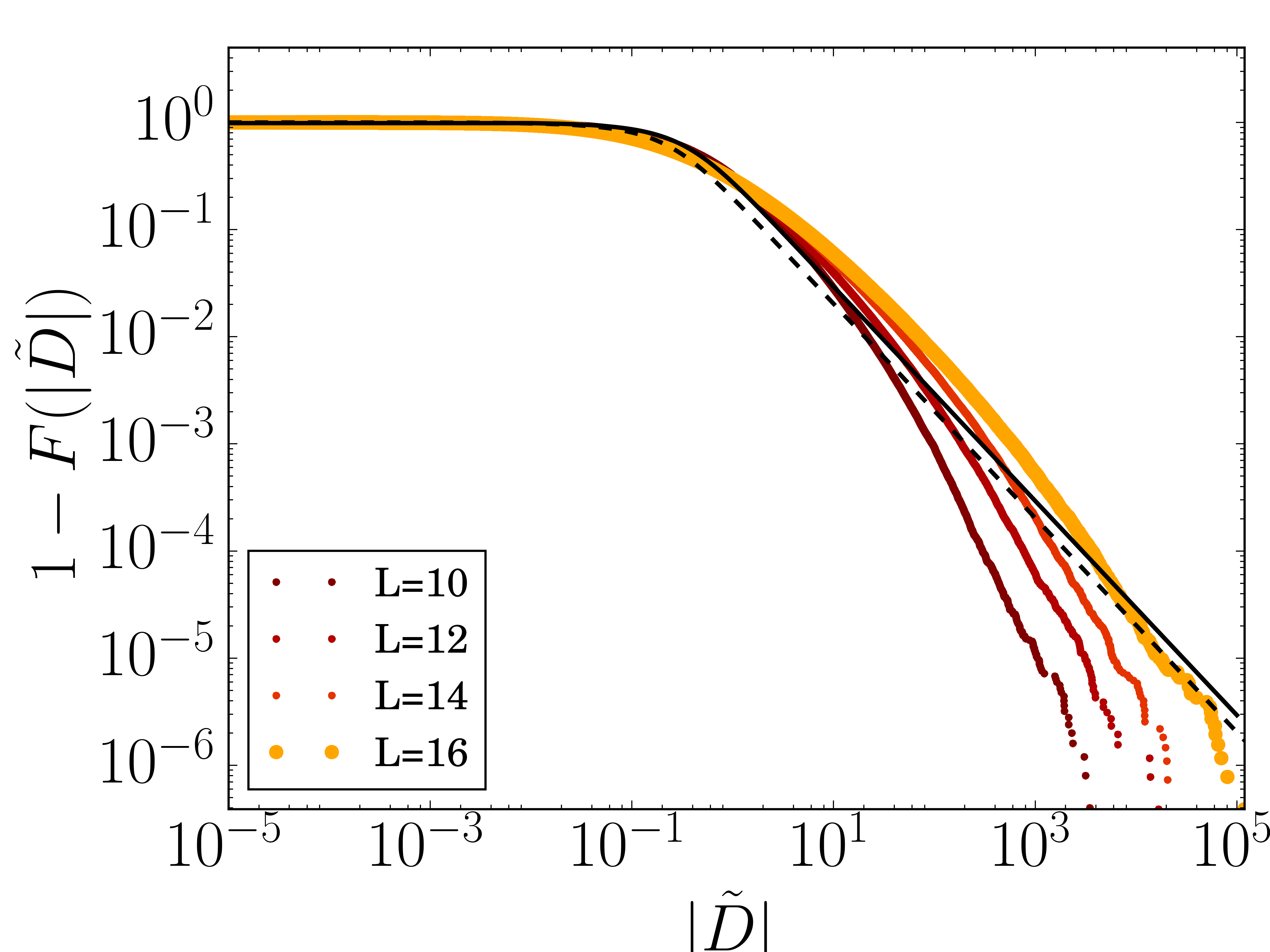}
\includegraphics[width=.45\textwidth]{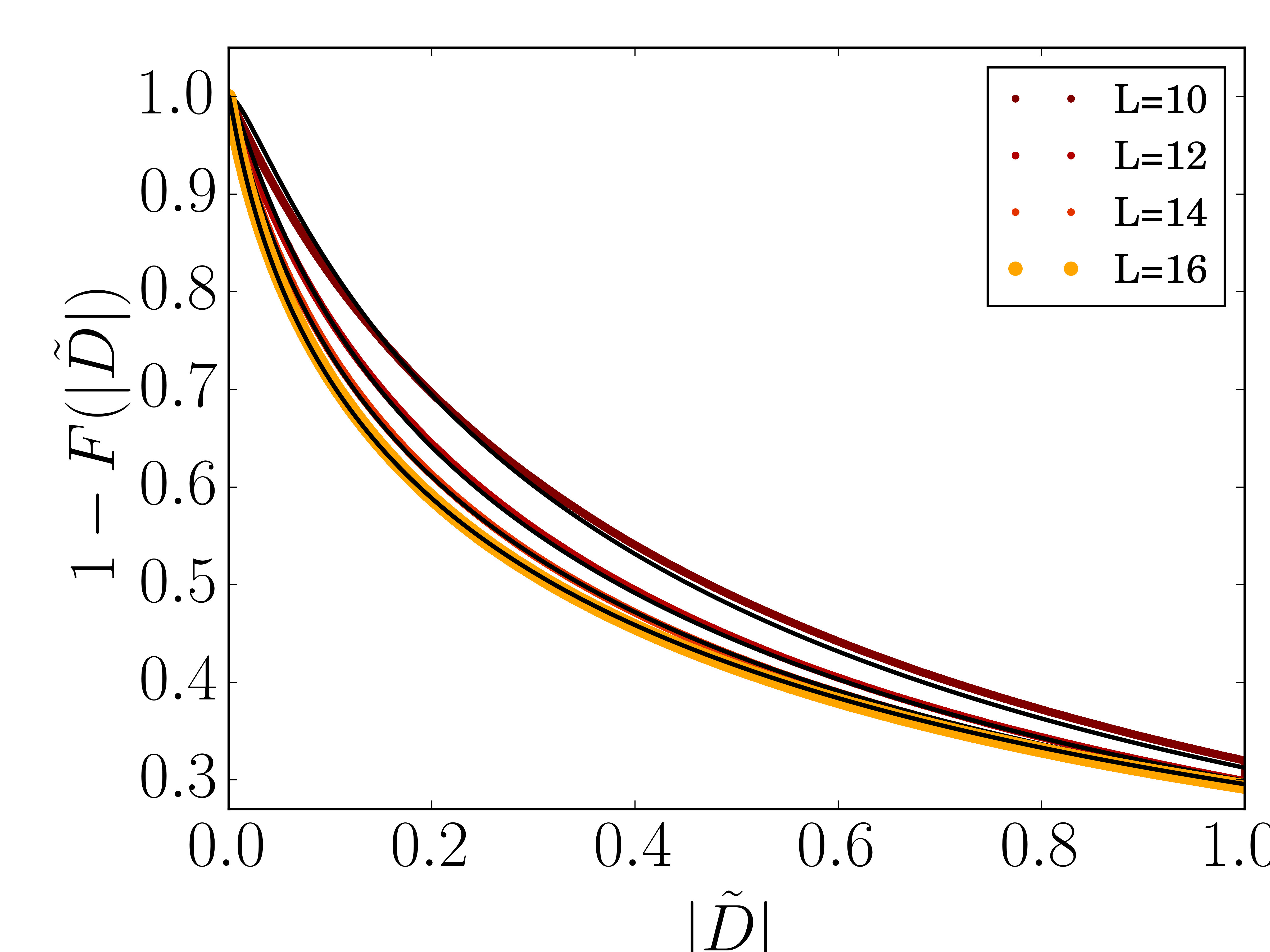}
\caption{Same as in Fig. \ref{fig:sizermt}, but in the localized regime ($W/t=5.5$).  The left panel is in log scale and shows system-size dependence of the tails of the Drude weight distribution, which is not described by the Cauchy distribution (solid line for $L=10$ and dashed line for $L=16$). The right panel is in linear scale and shows the bulk of the distribution. The curves are not described by the Cauchy distribution as well (see main text), but are in good agreement, for all system sizes, with a
log-normal distribution (solid black lines). }\label{fig:sizeloc}
\end{center}
\end{figure}

\end{document}